\documentclass[final,3p,twocolumn,times]{elsarticle}

\usepackage{xcolor,soul}
\usepackage{graphicx}
\usepackage{comment}
\usepackage{subfig}
\usepackage{url}
\usepackage{cleveref}
\usepackage{multirow}
\usepackage{acronym}
\usepackage[inline]{enumitem}
\usepackage{booktabs}

\acrodef{AI}{Artificial Intelligence}
\acrodef{UE}{User Equipment}
\acrodef{MC}{MECPerf Client}
\acrodef{MO}{MECPerf Observer}
\acrodef{MRS}{MECPerf Remote Server}
\acrodef{MA}{MECPerf Aggregator}
\acrodef{QoE}{Quality of Experience}
\acrodef{API}{Application Programming Interface}
\acrodef{FSM}{Finite State Machine}
\acrodef{NAT}{Network Address Translation}
\acrodef{3GPP}{Third Generation Partnership Project}
\acrodef{CDF}{Cumulative Distribution Function}
\acrodef{5G-PPP}{5G Public Private Partnership}
\acrodef{AA}{Authentication and Authorization}
\acrodef{AP}{Access Point}
\acrodef{AR}{Augmented Reality}
\acrodef{BGP}{Border Gateway Protocol}
\acrodef{BS}{Base Station}
\acrodef{CFS}{Customer Facing Service}
\acrodef{CPU}{Central Processing Unit}
\acrodef{DB}{Database}
\acrodef{DNS}{Domain Name System}
\acrodef{ETSI}{European Telecommunications Standards Institute}
\acrodef{FCFS}{First Come First Serve}
\acrodef{FaaS}{Function as a Service}
\acrodef{NFaaS}{Named Function as a Service}
\acrodef{GSMA}{Global System for Mobile communications Association}
\acrodef{GPU}{Graphics Processing Unit}
\acrodef{HMI}{Human Machine Interface}
\acrodef{HTML}{HyperText Markup Language}
\acrodef{HTTP}{Hyper-Text Transfer Protocol}
\acrodef{ICN}{Information-Centric Networking}
\acrodef{IETF}{Internet Engineering Task Force}
\acrodef{IIoT}{Industrial Internet of Things}
\acrodef{IPP}{Interrupted Poisson Process}
\acrodef{IP}{Internet Protocol}
\acrodef{ISG}{Industry Specification Group}
\acrodef{ITS}{Intelligent Transportation System}
\acrodef{ITU}{International Telecommunication Union}
\acrodef{IT}{Information Technology}
\acrodef{IaaS}{Infrastructure as a Service}
\acrodef{IoT}{Internet of Things}
\acrodef{JSON}{JavaScript Object Notation}
\acrodef{LCM}{Life Cycle Management}
\acrodef{LL}{Link Layer}
\acrodef{LTE}{Long Term Evolution}
\acrodef{MAC}{Medium Access Layer}
\acrodef{MBWA}{Mobile Broadband Wireless Access}
\acrodef{MCC}{Mobile Cloud Computing}
\acrodef{MEC}{Multi-access Edge Computing}
\acrodef{MEPM}{MEC Platform Manager}
\acrodef{ML}{Machine Learning}
\acrodef{MNO}{Mobile Network Operator}
\acrodef{NFV}{Network Function Virtualization}
\acrodef{OSPF}{Open Shortest Path First}
\acrodef{OSS}{Operations Support System}
\acrodef{OS}{Operating System}
\acrodef{OWC}{OpenWhisk Controller}
\acrodef{OTT}{Over The Top}
\acrodef{PMF}{Probability Mass Function}
\acrodef{PoA}{Point of Attachment}
\acrodef{PU}{Processing Unit}
\acrodef{PaaS}{Platform as a Service}
\acrodef{QoE}{Quality of Experience}
\acrodef{QoS}{Quality of Service}
\acrodef{RPC}{Remote Procedure Call}
\acrodef{RR}{Round Robin}
\acrodef{RSU}{Road Side Unit}
\acrodef{SDN}{Software Defined Networking}
\acrodef{SRPT}{Shortest Remaining Processing Time}
\acrodef{SaaS}{Software as a Service}
\acrodef{TCP}{Transmission Control Protocol}
\acrodef{TLS}{Transport Layer Security}
\acrodef{TSN}{Time-Sensitive Networking}
\acrodef{UDP}{User Datagram Protocol}
\acrodef{UE}{User Equipment}
\acrodef{URI}{Uniform Resource Identifier}
\acrodef{URL}{Uniform Resource Locator}
\acrodef{UT}{User Terminal}
\acrodef{VANET}{Vehicular Ad-hoc Network}
\acrodef{VIM}{Virtual Infrastructure Manager}
\acrodef{VM}{Virtual Machine}
\acrodef{VNF}{Virtual Network Function}
\acrodef{VR}{Virtual Reality}
\acrodef{WLAN}{Wireless Local Area Network}
\acrodef{WRR}{Weighted Round Robin}
\acrodef{YAML}{YAML Ain't Markup Language}
\acrodef{STL}{Standard Template Library}

\usepackage{ulem}

\journal{Computer Networks}

\begin{document}

\begin{frontmatter}
\title{Measurement-Driven Design and Runtime Optimization in Edge Computing:\\Methodology and Tools}

\author[unifi,dip]{Chiara Caiazza}
\ead{chiara.caiazza@unifi.it}
\author[iit]{Claudio~Cicconetti}
\ead{claudio.cicconetti@iit.cnr.it}
\author[iit]{Valerio~Luconi}
\ead{valerio.luconi@iit.cnr.it}
\author[dip]{Alessio~Vecchio\corref{cor1}}
\ead{alessio.vecchio@unipi.it}

\cortext[cor1]{Corresponding author}
\address[dip]{Dip. di Ing. dell'Informazione, Universit\`a di Pisa, Largo L. Lazzarino 1, 56122 Pisa, Italy}
\address[unifi]{University of Florence, Italy}
\address[iit]{Istituto di Informatica e Telematica, Consiglio Nazionale delle Ricerche, Via G. Moruzzi, 1, 56124 Pisa, Italy}

\date{January 2021}

\begin{abstract}
Edge computing is projected to become the dominant form of cloud computing in the future because of the significant advantages it brings to both users (less latency, higher throughput) and telecom operators (less Internet traffic, more local management).
However, to fully unlock its potential at scale, system designers and automated optimization systems alike will have to monitor closely the dynamics of both processing and communication facilities.
Especially the latter is often neglected in current systems since network performance in cloud computing plays only a minor role.
In this paper, we propose the architecture of MECPerf, which is a solution to collect network measurements in a live edge computing domain, to be collected for offline provisioning analysis and simulations, or to be provided in real-time for on-line system optimization.
MECPerf has been validated in a realistic testbed funded by the European Commission (Fed4Fire+), and we describe here a summary of the results, which are fully available as open data and through a Python library to expedite their utilization.
This is demonstrated via a use case involving the optimization of a system parameter for migrating clients in a federated edge computing system adopting the GSMA platform operator concept.
\end{abstract}

\begin{keyword}
Edge computing, ETSI MEC, network measurements, GSMA platform operator
\end{keyword}

\end{frontmatter}

\section{Introduction}
\label{sec:introduction}

The widespread adoption of edge computing is shifting the computation from remote data centers to locations closer to the users~\cite{Campbell2019} for mainly three reasons:
i) users in some communities (e.g., Industry 4.0~\cite{9037362}) or for some
applications (e.g., real-time face recognition or behavior
analysis~\cite{Wang2017a}) may have concerns regarding the privacy of their
data being processed in the public cloud;
ii) reaching remote data centers incurs an extra delay, which
latency-sensitive applications (e.g., \ac{AR}/\ac{VR}~\cite{Braud2017}) might
be unwilling to pay;
iii) some applications have very high bandwidth requirements (e.g., real-time
video analytics~\cite{8567664}), which increases the cost of outbound traffic
and may cause congestion on the network links towards the Internet.
Edge computing will result in many applications with a very high social and economic impact, like healthcare IoT~\cite{8950450}, multimedia IoT~\cite{8993839} and Industry 4.0~\cite{krupitzer2020survey}, to be unlocked at their full potential, as currently unattainable thanks to: ultra low latency, very high throughput, indoor/outdoor geographical localization, on-premises processing of \ac{AI}/\ac{ML} data streams.

In the scientific literature and the market press, the term \textit{edge computing} is associated with many different concepts, which we can classify into three broad categories, depending on where the services run: i) on \textit{end devices} sharing their available computation capabilities with peers (e.g., \cite{8662815}); ii) on \textit{far-edge} devices that are very close to the end devices, typically just one hop away like \ac{IoT} gateways~\cite{Bellavista2019} or base stations in cellular systems~\cite{Guo2018a}.; and, iii) on \textit{near-edge} devices, which have been provisioned in the core network of telecom operators for the purpose of providing a better service for some applications~\cite{Liu2020}.
In this paper, we focus on the latter, i.e., which is being covered by all major cloud providers in partnership with telecom operators (e.g., Microsoft Azure Zones, AWS Wavelength) and is the target of the \ac{ETSI} \ac{MEC} industry study group~\cite{Filippou2020}, which has released at the time of writing a reference architecture 2.0 and related \acp{API} to take advantage of emerging communication paradigms, such as in a connected-vehicle scenario~\cite{8515152}.

However, transitioning from cloud computing to edge computing will not be fully transparent to either platforms and applications, and in fact, there are several aspects that are still under scrutiny by the research community~\cite{Ramachandran2021}.
In this work, we focus on the collection of \textit{network performance measurements} and on their use for improving design and runtime decisions.
Network measurements are relevant because the network alone is responsible for two of the three major reasons why one service provider can be interested in edge computing, i.e., \textit{delay} and \textit{bandwidth} (the other one is \textit{privacy}, which we do not consider in this paper).
In all the scenarios where there is high variability of the network conditions,
which can be due to the individual devices (especially if mobile), to the
aggregate traffic of a group of user devices, or to the backhaul connections
(i.e., the links between the core network
of the operator and the Internet), it is of the utmost importance to
monitor the network performance and take appropriate decisions: for a user
device this can mean migrating to another network or modifying some
application-level parameters~\cite{9040261}; on the other hand,
the platform might allocate further resources, e.g., via~\ac{SDN} and~\ac{NFV}
solutions~\cite{Pan2017}.

The objective of this paper is many-fold:

\begin{enumerate}[parsep=0em,leftmargin=*,label=\arabic*.]
\item illustrate the motivation of MECPerf as a toolbox for collecting and using measurements in an edge system with the goal of runtime optimization at application-, service-, or system-level (Section~\ref{sec:datacollection});

\item describe the architecture of the \textit{MECPerf Collection System}, which is designed for the acquisition of measurements in three modes (active, passive, self), each with a different objective and performance implications (Section~\ref{sec:mecperfcol});

\item describe the \textit{MECPerf Library}, which we have developed to allow researchers and practitioners to easily access our open dataset and integrate it in their own performance evaluation tools (Section~\ref{sec:mecperflib});

\item disseminate the results we have obtained in experiments carried out within the Fed4Fire+ project\footnote{Funded by the European Commission in the H2020 program, with grant no. 732638, \url{https://www.fed4fire.eu/}.}, which consisted of measuring the network performance in a realistic edge computing testbed, with Wi-Fi and LTE access technologies (Section~\ref{sec:experiments});

\item discuss a case study where orchestration extends across two federated edge computing infrastructures as an illustrative example of the MECPerf collection system and library (Section~\ref{sec:casestudy}).
\end{enumerate}

Before delving on the main contribution sections of the paper, we summarize the most relevant literature on this topic in Section~\ref{sec:relwork}, while concluding remarks are reported in Section~\ref{sec:conclusion}.
All software tools, both for collecting and retrieving the data, are made available as \textit{open source} on GitHub\footnote{\url{https://github.com/MECPerf}}, and the full dataset is \textit{open data} and can be retrieved from Zenodo\footnote{\url{https://zenodo.org/record/4647753#.YGNLDnUzZH4}}.
\section{Related Work}
\label{sec:relwork}


Even though edge computing is a relatively recent technology trend, it has
already attracted significant interest in the scientific and industrial
research communities.
As a matter of fact, there are already survey papers dealing with specific
research aspects: computation offloading modeling~\cite{Lin2020}, resource
management~\cite{Hong2019}, communication aspects~\cite{8391395},
service orchestration~\cite{Taleb2017a}.
The interested reader can also find a 360-degree introduction
in~\cite{Bellavista2019}.

\subsection{Open Data on Edge Computing}

Few works had, as one of their main goals, the collection of open performance data about edge computing systems.
A collection of experimental data was carried out on a prototype testbed implementation of a Mixed Reality (MR) application based on edge computing, to be the basis for further simulative studies~\cite{tocze2020characterization}. Edge workload traces derived from speech-based and MR applications are publicly available \cite{klervie_tocze_2020_3974220}. 
Another available dataset provides measurements about placement and execution of functions in a mixed cloud-edge cluster~\cite{rashed_alexander_2019_3527672}. Some summary statistics about an edge-based architecture, where the computing facilities are pushed in the proximity of, or co-located on, the machine that provides the radio access services, are also available~\cite{apostolaras_apostolis_2020_3731140}.
The execution traces of an ML application were first acquired on a real testbed and then used for evaluating the container scheduling strategies for serverless edge computing~\cite{RAUSCH2021259}. Traces are publicly available and contain the execution time of functions belonging to the considered machine learning workflow, as well as the time needed for transferring the data~\cite{alexander_rashed_2020_3628454}.
Latency traces between six cloud data centers have been collected in a study aimed at improving the performance of state machine replication \cite{10.1145/3386367.3431291}. The traces are available to the public but do not include edge-based measurements.
A different approach is proposed in \cite{10.1145/3419394.3423643}, where a workflow called DoppelGANger, based on generative adversarial networks, is used to build open synthetic datasets with minimal expert knowledge. DoppelGANger is specifically designed to produce time-series belonging to the networking domain (e.g., loss rate) together with the associated metadata (e.g., ISP).

Overall, these studies highlight that the availability of data collected from real systems is of paramount importance in the evaluation of applications executed in the edge.
However, no comprehensive dataset encompassing all the key characteristics of edge computing is available today~\cite{Kolosov2020} and the
available data generally focus on the computing perspective of edge infrastructure, whereas the networking performance is only marginally considered.

\subsection{Benchmarking Edge Computing Platforms}

EdgeBench is a benchmarking solution tailored for serverless edge computing, based on the execution of some canonical applications~\cite{Das2019}. In particular, EdgeBench was used to compare the performance of two commercial edge/cloud providers (AWS Greengrass and Azure IoT Edge). 
Another study analyzed three connectivity options between large cloud providers~\cite{yeganeh2020first}. The three options, based on the public Internet, private Cloud connectivity, or third-party private providers, were characterized using active measurements (but without a specific focus on the edge).
DeFog is a benchmarking suite aimed at evaluating the performance of a fog-based platform compared to a cloud-based one~\cite{McChesney2019}. Benchmarking relies on the execution of a number of containerized applications. Applications include speech-to-text conversion, real-time face detection from video streams, and others that are suitable for being executed in cloud-only and fog-only mode.  A mixed cloud-fog mode that involves long-haul communication between a central cloud infrastructure and peripheral nodes is also considered. Collected metrics include latency, number of CPUs/cores, execution time, amount of bytes transferred, and several others. In~\cite{McChesney2019}, the authors highlighted the lack of publicly available benchmarks useful to estimate the advantages of fog/edge-based computing compared to more traditional solutions. We share the same motivation, but our approach takes different directions compared to DeFog. First, the metrics we collect are more network-oriented. Second, our goal is not really related to benchmarking: we do not compute summary indexes but collect raw data that can be used for optimizing runtime decisions or improve the design of edge systems and applications.

\subsection{Edge Computing Emulators and Simulators}
\label{subsec:sim}

\begin{table*}[!t]
    \footnotesize
    \centering
    \begin{tabular}{p{0.10\textwidth} p{0.08\textwidth} p{0.06\textwidth} p{0.17\textwidth}
    p{0.51\textwidth}}
        \toprule
        \textbf{Tool} & \textbf{Based on} & \textbf{Type} & \textbf{Simulated/emulated devices} & \textbf{Network model} \\
        \midrule
        
        openLEON \cite{Fiandrino2019} & 
        srsLTE, mininet & 
        Emulator& 
        Edge data center, and cloud data center  &  
        \textbf{End-devices access technology:} LTE.\newline
        \textbf{Network topology:} a real smartphone connects to the eNB using dedicated hardware. The rest of the network is emulated using Containernet. \newline
        \textbf{Link characteristics:} real, between user equipment and eNB, emulated according to mininet the rest. \newline
        \textbf{Network stack:} real.\\
        &&&\\
        
        SPHERE \cite{FERNANDEZCERERO2020101966} & 
        SCORE simulator & 
        Simulator&
        IoT, mobile devices, independent clusters of cloudlets and clouds& 
        \textbf{End-devices access technology:} not simulated. \newline
        \textbf{Network topology:} defined by the programmer.\newline
        \textbf{Link characteristics:} deterministic latency and bandwidth. \newline
        \textbf{Network stack:} not simulated.\\
        &&&\\

        FogNetSim++ \cite{Qayyum2018}&
        OMNeT++, INET&
        Simulator&
        IoT, mobile devices, fog nodes, and cloud data center& 
        \textbf{End-devices access technology:} both wired and wireless. \newline
        \textbf{Network topology:} defined through a GUI or a configuration file.\newline
        \textbf{Link characteristics:} according to INET models. \newline
        \textbf{Network stack:}  simulated physical, link-layer, network, transport, and application layers. \\
        &&&\\
               
        iFogSim \cite{https://doi.org/10.1002/spe.2509}& 
        CloudSim&
        Simulator&
        IoT, fog nodes, cloud data center&
        \textbf{End-devices access technology:} wireless \newline
        \textbf{Network topology:} defined by through a GUI or a configuration file.\newline
        \textbf{Link characteristics:} fixed latency and bandwidth.\newline
        \textbf{Network stack:} not simulated.\\
        &&&\\

        EdgeCloudSim \cite{10.1002/ett.3493}&
        CloudSim&
        Simulator&
        Mobile devices, edge servers, and cloud data center&
        \textbf{End-devices access technology:} wireless. \newline
        \textbf{Network topology:} defined through an XML file.\newline
        \textbf{Link characteristics:} simple queue model, empirically derived properties. \newline
        \textbf{Network stack:} not simulated. \\
        &&&\\

        YAFS \cite{Lera2019a}&
        Python, Simpy, and NetworkX&
        Simulator&
        IoT, fog nodes, and cloud data center&
        \textbf{End-devices access technology:} not simulated.\newline 
        \textbf{Network topology:} defined through a configuration file or imported (CAIDA, BRITE topologies).\newline
        \textbf{Link characteristics:} fixed latency and bandwidth.\newline
        \textbf{Network stack:} not simulated.\\
        &&&\\

        ECSim++ \cite{Nguyen2018}&
        OMNeT++, INET&
        Simulator&
        Mobile devices, edge nodes, and clouds &
        \textbf{End-devices access technology:} both wired and wireless. \newline
        \textbf{Network topology:} defined through a GUI or a configuration file.\newline
        \textbf{Link characteristics:} according to INET models.\newline
        \textbf{Network stack:} simulated physical, link-layer, network, transport, and application layers.\\
        \bottomrule
    \end{tabular}
    \caption{A comparison of the main network-related characteristics of edge computing emulators and simulators.}
    \label{Table:edgeemulatorandsimulatorcomparison}
\end{table*}

Emulators can be used for testing real applications and protocols under network conditions defined by the experimenter, whereas simulators are used for evaluating the impact of design choices and tuning the parameters of operations.

openLEON is an end-to-end emulation platform that covers the network between a mobile user and the edge data center~\cite{Fiandrino2019}. The system is based on srsLTE~\cite{srslte} for the wireless part and Containernet~\cite{peuster2016medicine} for the emulation of the data center. Being an emulation platform that includes some hardware elements, the path followed by openLEON is clearly different from our one, which is instead dedicated to software-only solutions. 

There are several simulators focusing on the different facets of edge computing. A simulator about centralized, distributed, and hybrid orchestration models for clusters of edge nodes is described in~\cite{FERNANDEZCERERO2020101966}. In particular, each cluster can be turned on/off depending on network-/cluster-level efficiency and performance strategies.
FogNetSim++ is a simulator based on OMNeT++ aimed at evaluating task scheduling algorithms and considering factors such as the utilization of fog nodes, SLAs, and handovers~\cite{Qayyum2018}.
iFogSim, instead, allows evaluating the performance of \ac{IoT} applications in terms of latency, network congestion, power consumption, and other cost metrics~\cite{https://doi.org/10.1002/spe.2509}. iFogSim provides different placement strategies to deploy application modules in the edge-cloud continuum. 
EdgeCloudSim is a simulator based on CloudSim aimed at studying orchestration of VMs, resource management, and task offloading~\cite{10.1002/ett.3493}. 
YAFS is a simulator for IoT scenarios in fog computing that allows defining complex networks by importing CAIDA and BRITE topologies, but where links are characterized by deterministic bandwidth and latency values~\cite{Lera2019a}.

Table \ref{Table:edgeemulatorandsimulatorcomparison} summarizes the main characteristics of the above emulation and simulation platforms.
For the vast majority of them, the network model does not include the transport, network, data-link, and physical layers, and communication between the considered end-points is just characterized according to simple bandwidth and delay models.
The adoption of simple network models is probably due to considering edge computing just a variation of cloud computing (many of the above-mentioned simulators are derived from already existing cloud-specific ones). However, while in a cloud-centric scenario, the assumption of links characterized by fixed values of bandwidth and delay is reasonable, the same cannot be said when including the elements at the edge of the network: the presence of wireless links make connections possibly intermittent and characterized by significant changes in terms of observed performance. Similar considerations can be made about long-haul paths connecting the periphery of the network to a centralized cloud infrastructure.

Some simulators, such as ECSim++ \cite{Nguyen2018}, include a more detailed model of the network, with support for simulating the transport layer and the lower ones. However, some parts of a typical edge-cloud infrastructure cannot be easily modeled, for instance, the backhaul connections and the presence of cross-traffic. 
MECPerf allows experimenters the collection of traces in a real setting and their use in simulators. As a consequence, MECPerf better supports the design of edge-based mechanisms where the interaction with a more realistic network model is pivotal.

\section{MECPerf Overview}
\label{sec:datacollection}

\begin{figure}[!t]
    \centering
    \includegraphics[width=0.47\textwidth]{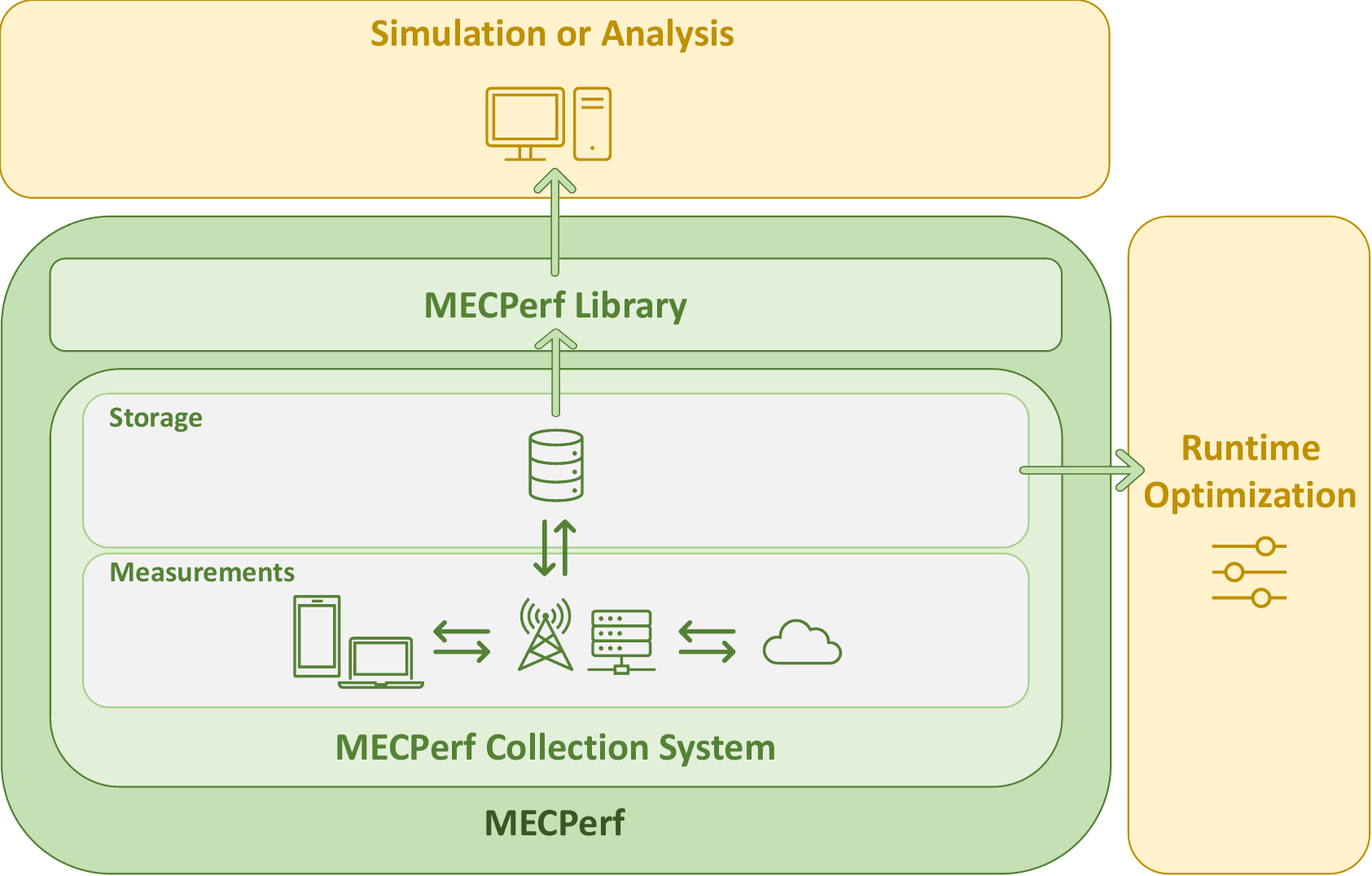}
    \caption{MECPerf Architecture. MECPerf is composed by two building blocks: i) the MECPerf Collection System, which runs measurements in the MEC and stores them for future use; ii) the MECPerf Library which is used to retrieve measurement traces that can be used for simulations or analyses}.
    \label{fig:overview}
\end{figure}

Design and runtime optimization in edge computing may benefit from the adoption of a \textit{measurement-driven methodology}, where network-related KPIs are collected and used to improve existing applications, services and systems. This can be done both at the application/service level, by supporting decisions (e.g., about migrating application components from the cloud to the edge), and at the system level, by providing insights about the performance of specific subsystems (e.g., the impact of different access technologies).

We identify the following three phases:

\begin{itemize}[parsep=0em,leftmargin=*]

\item \textit{Collection of KPIs.} Several flavors of delay and throughput can be collected along with the most relevant segments of the edge-cloud continuum and access network. Different approaches are possible to cope with the constraints imposed by the specific system under observation or the required level of integration. In many cases, the current status of the network is not enough when considered in isolation, and it must be combined/compared with information about the past dynamics. Thus, measurement results are saved onto persistent storage together with the main characteristics of the evaluated scenario to ease the extraction of information according to simple criteria.

\item \textit{Using KPIs in design.} A quantitative assessment of the network is the substrate upon which design choices can be taken.
Measurement results allow the designer to better understand the impact of the possible alternatives in terms of technologies and architectures. For instance, a simulator fed with the KPIs collected in a given scenario can make more clear if and how a given application must be decomposed for being relocated, as a whole or just in part, to the edge.

\item \textit{Using KPIs for runtime optimization.}  
Information about the current environment, in terms of network performance, can be used, at runtime, to optimize algorithms operating at the application/service level. 
\end{itemize}

For this reason, we designed and implemented MECPerf: a system for collecting network performance in an edge computing scenario and making the data easily available.  MECPerf is composed of two major elements: the MECPerf Collection System and the MECPerf Library (shown in Figure~\ref{fig:overview}). The former collects information about the delay and bandwidth of the most relevant network segments involved in a typical MEC-enabled application, such as the one between the users and the MEC node and the one between the MEC node and a centralized cloud. To collect network performance indicators, MECPerf relies on the presence of measuring elements deployed on users' devices, on MEC nodes, and on the other hosts that are relevant for the considered application, even when they are far from the edge of the network. The MECPerf Library allows experimenters to easily extract network performance traces that have been previously collected. In particular, experimenters just have to provide a high-level description of the properties of the network they are interested in to obtain the relevant data. The overall goal is to capture the real behavior of an edge-based network and use it, for instance, to improve the fidelity of simulative studies, which are frequently based on simple models for what concerns the characteristics of the network.

\section{The MECPerf Collection System}
\label{sec:mecperfcol}

In this section, we describe the architecture of the MECPerf Collection System, along with the measurement methods and collected metrics.

\subsection {Architecture}

\begin{figure}[!t]
    \centering
    \includegraphics[width=0.47\textwidth]{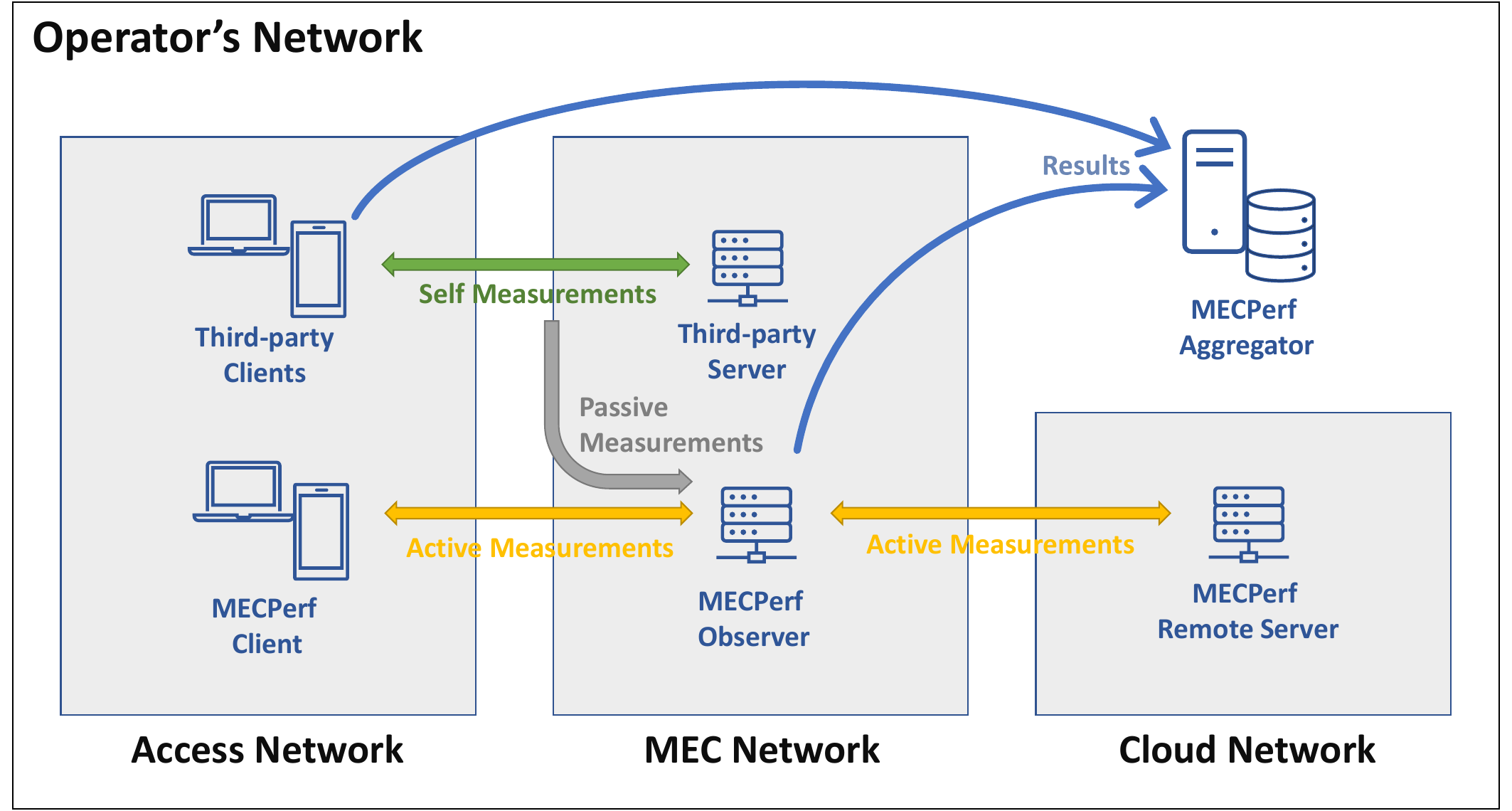}
    \caption{Architecture of the MECPerf Collection System.}
    \label{fig:architecture}
\end{figure}

Figure~\ref{fig:architecture} shows the architecture of the MECPerf Collection System, which is deployed inside a mobile network operator's network. The MECPerf Collection System consists of multiple components: i) the MECPerf Client, which runs on a UE in the access network; ii) the MECPerf Observer, which runs on MEC application servers in the MEC network; iii) the MECPerf Remote Server, which is executed on a centralized cloud; iv) the MECPerf Aggregator that can reside anywhere in the operator's network.

The MECPerf Client, the MECPerf Observer, and the MECPerf Remote Server cooperate to collect network KPIs. Three types of measurements are supported: active, self, and passive. In active measurements, pairs of MECPerf measuring end-points inject traffic onto the network to compute network performance indexes. Self measurements provide information about the performance of third-party applications. In particular, the application under observation is instrumented with code useful to contribute its measurements to the MECPerf repository. However, the production of traffic still takes place under the application's responsibility, i.e., the application does not generate additional traffic compared to the traffic it would generate in normal operations. In passive measurements, the traffic of applications is passively collected and analyzed to compute the metrics of interest. Differently from self measurements, applications are now unaware of the measuring process. Most of operations are, in passive measurements, carried out by the MECPerf Observer, as it is located on a particularly significant vantage point.
Finally, the MECPerf Aggregator stores the measures collected from the other components onto persistent memory, using a relational database, and makes the results available for future use. All the interactions with the MECPerf Aggregator, i.e., sending measurement results or extracting the collected metrics, are performed via a REST interface.

\subsection {Measurement Methods and Collected Metrics}

Different versions of delay and bandwidth metrics can be collected, depending on the needs of the experimenter and the characteristics of the system under observation. 
In general, delay and bandwidth, collected at different layers of the TCP/IP stack, are the most common metrics used to evaluate the performance of a network or networked applications. More complex (and application-specific) metrics can be derived from delay and bandwidth measurements (jitter or QoS can be typical examples). To keep the MECPerf Collection System simple and independent from specific applications, we decided to collect just these two metrics. However, the MECPerf Collection System is easily extensible, and more complex or specific metrics can be added in the future. In addition, as highlighted in the following paragraphs, application developers can provide the metrics that are most relevant for their application, enabling maximum flexibility in applications performance analysis.

As stated above, all the collected metrics are sent to the MECPerf Aggregator via a REST interface. The transfer is based on JSON objects that contain not only the measurement results but also the metadata useful for retrieving traces at a later time.

\subsubsection*{Active Measurements}
\label{subsubsection:measurementsMethods_activeMeasurement}

The metrics measurable through active methods are TCP bandwidth, UDP bottleneck capacity, and both TCP-based and UDP-based latency. The TCP bandwidth is computed at the receiver as the bandwidth of a data stream transfer, thus it is at the application layer, while the UDP bottleneck capacity is computed using a known technique based on UDP packets that are sent back to back~\cite{Prasad03:bandwidth} (other, more sophisticated, techniques can also reuse the packets normally generated by applications to reduce the amount of injected traffic~\cite{DBLP:conf/pam/AhmedMS20}). The bottleneck capacity is the capacity of the narrowest link in a path and it is computed at the network layer. Capacity is not dependent on the presence of cross-traffic, at least in principle, as it represents a physical property of the link. Finally, the latency between two end-points is computed by sending a message and waiting for the response, then computing the round-trip time (RTT). Both UDP and TCP can be used when measuring latency. TCP latency is an application layer measurement, as TCP is equipped with connection-oriented and reliability mechanisms. UDP is instead best effort, thus the UDP latency measurement can be considered at the network/transport layer. Each metric is collected in all the network segments of the MECPerf Collection System's architecture, as highlighted in Figure~\ref{fig:architecture}, i.e., in the access-MEC segment between MECPerf Client and MECPerf Observer, and in the MEC-cloud segment between MECPerf Observer and MECPerf Remote Server. Each metric can be taken in both directions, i.e., upstream and downstream. Finally, all metrics are sent by the MECPerf Observer to the MECPerf Aggregator, via the REST interface.

The metrics collected with active measurements are intended to characterize the general performance of the MEC and cloud infrastructure of a mobile network operator independently from specific applications. The results obtained with these measurements can enable the evaluation of the suitability of a specific MEC/cloud configuration for the desired application. For example, for applications like 4K or 8K video streaming, or 360-degree video, the bottleneck capacity measurements could be useful to preliminary understand if the network edge infrastructure will be able to handle their massive bandwidth requirements. Active latency measurements will be useful to developers of applications such as autonomous vehicle or traffic management in a smart city context, as they need to know if the network will be able to provide the extremely low latency requirements needed for those applications. In addition, having measurements for different segments of the network infrastructure will be able to help application developers decouple their application's functions in the most efficient way.

\begin{figure}[t!]
    \centering
    \includegraphics[width=0.36\textwidth]{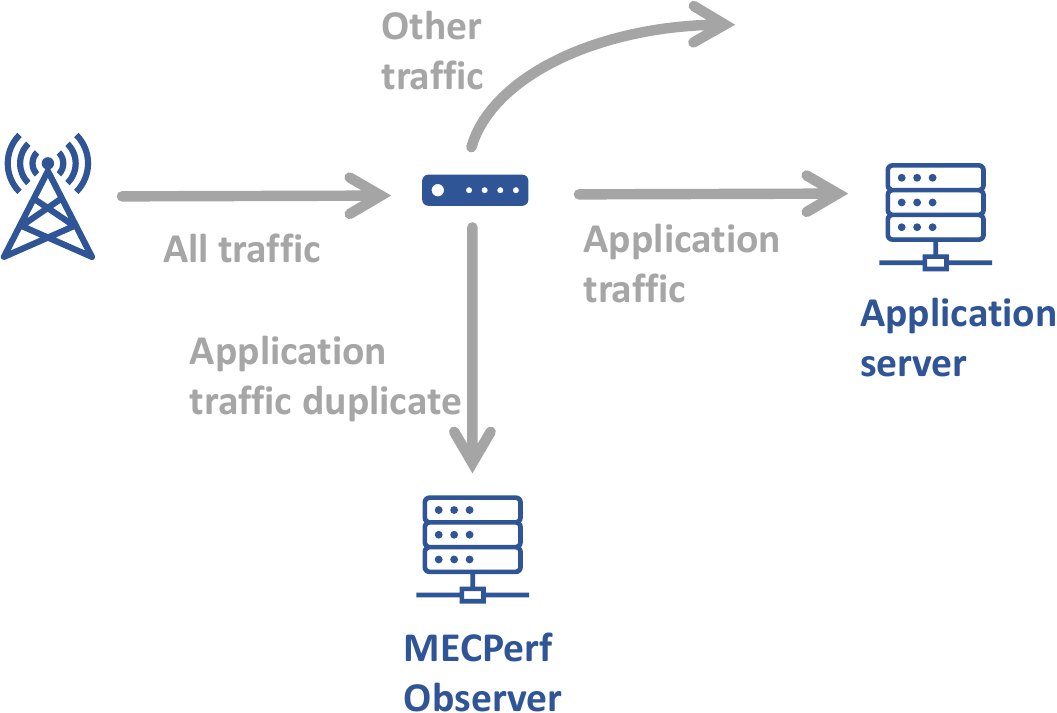}
    \caption{The traffic redirection mechanism for the passive measurements through traffic analysis.}
    \label{fig:passiveMeasurements_traffic-redirection}
\end{figure}

\subsubsection*{Self Measurements}
In this case, an application can choose when and how its performance indicators must be collected. This means that different client applications might measure different metrics, as every application may use a slightly different definition. For instance, a video streaming application may collect the average bitrate obtained during application-defined intervals. Then, the client application shares the collected metrics with the MECPerf Aggregator using the REST interface. 
This collection method is intended to give the possibility for different application developers or providers to collect their custom metrics according to the intrinsic characteristics and requirements of a specific application. The decision on which metrics to collect, how to collect them, and when to provide them to the MECPerf Collection System is completely up to the application developers, as well as the responsibility for the implementation of the desired measurement functionalities, which will require modifications to the original application. Self measurements are particularly useful as they provide application-aware feedback for both application providers and the network operator itself on how the applications are performing and which applications are experiencing performance degradation due to the network status. For example, an autonomous vehicle application, having information about the performance of its network functions, can improve its decision-making processes by choosing whether or not to rely on network assistance, and for example reduce speed or increase distance from other vehicles if the network is not performing well. Ultra HD video streaming applications could increase or decrease buffering capabilities according to the measured network performance in terms of bandwidth and latency. On the network side, having information about the performance experienced by all the applications running on the edge infrastructure, and knowing their requirements, can help runtime optimization of resources, such as adding/removing VMs or moving VMs from cloud to edge or vice-versa.

\subsubsection*{Passive Measurements}

Passive measurements rely on a traffic redirection mechanism, as shown in Figure \ref{fig:passiveMeasurements_traffic-redirection}. The data generated from the client application is transparently redirected to a packet sniffer that generates pcap files containing traffic traces. Then, the application traffic is sent to the application server while the pcap file is sent to the MECPerf Observer. When the MECPerf Observer receives a pcap file, it looks for packets that match the application under observation. For each flow, the MECPerf Observer uses the timestamp and the payload size of each packet to compute the throughput of both TCP and UDP traffic. The throughput is computed using the payload of the transport layer, i.e., the application layer data. Instead, latencies are computed as the difference between the time of arrival of the ACK packet and the correspondent ACKed packet previously sent. This means that latency metrics can be computed only for TCP flows, and are to be considered at the transport layer. After completing the analysis of the file, the MECPerf Observer uses the REST interface to send all the computed metrics to the MECPerf Aggregator. It must be noted that the performance indices computed in this way are those as observed from an external point of view, which is unaware of application dynamics. This means that if the application is just staying idle measurements cannot be collected.
For this reason, we choose to collect just the general metrics such as bandwidth and delay, which can be suitable for a basic performance evaluation of all networked applications. In particular, the applications whose performance is best captured by passive measurements are the ones that produce a constant stream of data, such as file transfer. Other kinds of applications that produce intermittent traffic patterns, as for example, vehicle to vehicle communication, are not best characterized with this measurement approach, as it lacks the knowledge of the application internal functioning. However, in absence of self measurements, the passive measurements collection can still be valuable, as it allows an at least rough analysis of the performance of applications that do not share their metrics.

\subsubsection*{Performance Implications}

Active measurements are intended to be performed in a condition of an unloaded network, thus they do not interfere with the overall system performance. Passive measurements are computed by parsing pcap files collected via a traffic sniffing mechanism. To not interfere with the system's performance, there is no need to parse the files in a hard real-time fashion. The system can thus be tuned to run this task at low priority for not degrading the overall performance. In addition, the analysis of the pcap files can be performed on a dedicated infrastructure (machines and networks) separated from the production environment. Self measurements are completely under the control of application developers, and it is up to them to implement these measurements to not degrade their application's performance. Finally, it must be noted that sending the measurements to the MECPerf Aggregator uses network resources, even if these data transfers are typically in the order of few kilobytes per second or even less, thus a negligible amount in modern networks.

\subsection{Validation}

The implementation of the active mechanisms for measuring the delay and the bandwidth has been validated using a controlled testbed. In particular, artificial latency values and bandwidth restrictions were applied using tc-netem, and we verified that the measured values were consistent with the set ones (the details can be found in~\cite{9202841}).
Passive measurements have been validated in a similar way. We issued a TCP data transfer with iPerf between two machines hosted on the same LAN at the University of Pisa, and we applied bandwidth and latency restrictions with tc-netem. In particular, we restricted the bandwidth to 10, 20, 30, 40, and 50 Mbps and set additional latencies to 10, 50, and 100 ms. In all cases, the bandwidth and latency values measured with passive measurements were consistent with the set ones. In addition, the measured bandwidth values were consistent with the ones measured with iPerf. In particular, for the bandwidth restrictions, MECPerf passive measurements were able to measure 9.6, 19.1, 28.7, 38.2, and 47.4 Mbps, which differ by less than 1\% from the values measured by iPerf. The measured latencies were instead 10.8, 50.8, and 100.8 ms, being 0.8 ms the latency between the two machines when tc-netem is disabled.
Self measurements do not need any validation as they are under the application responsibility.
\section{The MECPerf Library}
\label{sec:mecperflib}

The measuring tools previously described are coupled with a library that eases the implementation of simulators based on real-world network conditions. In detail, the library provides access to trace-based bandwidth and latency values collected in a real setting, and it can be included in simulators aimed at understanding the behavior of applications and protocols when operating in an edge computing environment.

\subsection{Architecture}

\begin{figure}[t!]
    \centering
     \includegraphics[width=0.48\textwidth]{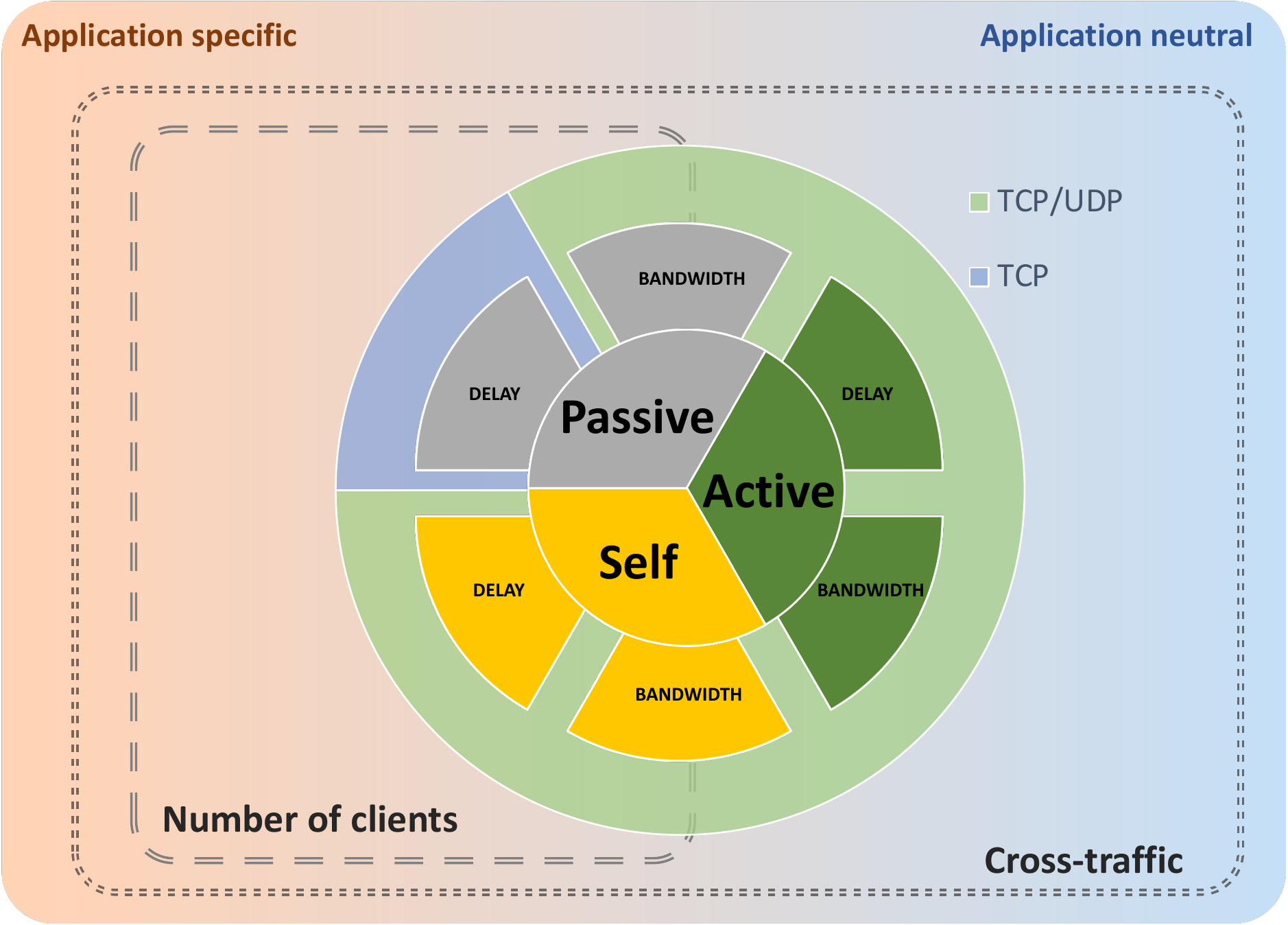}
    \caption{Type of measurements and related properties. Collected measurements are available for the three approaches, both in terms of delay and bandwidth. For each metric, both UDP- and TCP-based versions are available, with the exception of the passive-delay metric which is available only in the TCP-based version. For all of them, there are multiple levels of cross-traffic and number of clients (the latter only for passive and self approaches). Active measurements are more application neutral, whereas self and passive ones are more application specific.
    \label{fig:concept}}
\end{figure}

A simulator that wants to use the network metrics collected by MECPerf mainly interacts with a NetworkTraceManager (NTM) entity. The NTM shields the simulator from the intricacies of accessing the metrics stored in raw files. In particular, it allows the programmer to express, by means of a configuration descriptor, the network properties of the environment under study.
The library is composed of the NTM class and a set of input files containing bandwidth and latency measures obtained from real-world experiments. For each experiment, there is an input file containing the raw observed values and their timestamps. The NTM class uses the input files to extract bandwidth and RTT traces for a specific network segment. Then, whenever requested by the simulator logic, bandwidth and delay values are returned.
The library includes a JSON file that maps each input file with the setup used during the measure. The setup includes the type of measure, the segment measured, the direction of communication, the access technology used, and the amount of cross-traffic injected into the network. The properties that can be specified are visually represented in Figure \ref{fig:concept}.

The library is implemented in Python, and it is available at {\tt https://github.com/MECPerf}. The library is bundled together with the input files resulting from the experiments carried out in a testbed belonging to the Fed4FIRE+ infrastructure, according to the setup and procedures described in Section \ref{sec:experiments}. 

\subsection{Programming Interface}

When a simulator is interested in obtaining bandwidth or latency values for a network segment, it can instantiate an NTM object. First, the NTM receives as input a JSON description of the setup of interest and the seeds needed to ensure repeatable results. Then, the NTM uses the JSON description to identify the set of input files that correspond to the properties specified by the experimenter. Finally, the NTM randomly chooses an input file from the set of files matching the properties and uses it to generate the traces of latency and bandwidth values.

Once the traces are generated, bandwidth and latency values can be requested using the {\tt get\_bandwidth} and {\tt get\_rtt}  methods, which take as input the timestamp of the requested trace value. Since traces are collected in the real world, it is possible that there is no value corresponding to the specified timestamp, for example, because of a failed latency measurement. In these cases, NTM returns a value according to a sample-and-hold approach. It is also possible to manage a trace according to a circular logic, in case the required amount of samples exceeds the number of available ones. The library also takes care of aligning the bandwidth and latency traces when collected samples are not perfectly synchronized.

\section{Experiments and Collected Data}
\label{sec:experiments}

We collected experimental data using the NITOS testbed, hosted by the University of Thessaly, Greece. NITOS is a wireless experimentation facility that participates in the Fed4Fire+ European federation of Next-Generation Internet testbeds\footnote{https://www.fed4fire.eu/}. NITOS makes possible the execution of experiments using different wireless technologies (Wi-Fi, LTE) in a realistic setting. Table \ref{Table:testbedcharacteristics} summarizes the testbed characteristics. The setup also includes cloud infrastructure belonging to the University of Pisa, Italy. The two locations are connected through the public Internet to include long-haul connections in the experiments and thus take into account their impact when accessing resources hosted in a centralized cloud infrastructure (and in the end to highlight the possible differences between an edge-based solution compared to a more traditional one).

The purpose of these experiments is twofold. First, we show the capabilities of the MECPerf Collection System and how it can be used to obtain network and application performance indices that can help the optimization of network runtime operations. Second, we describe the most interesting characteristics of an extensive dataset of network measurements that we make available as \textit{open data} on Zenodo. The provided data is also used as input to the MECPerf Library and can be used to produce realistic simulations concerning an edge computing environment. In addition, we provide suggestions and useful takeaways on how the three measurement approaches can be combined to obtain a detailed picture of the performance of a MEC/cloud environment and the applications running in it.

\begin{table*}[!t]
    \small
        \begin{tabular}{cc}
            \begin{minipage}{.47\linewidth}
                \begin{tabular}{p{0.30\textwidth}|p{0.55\textwidth}}
                    \toprule
                    \multicolumn{2}{c}{\textbf{NITOS Indoor RF Isolated Testbed details}}\\ 
                    \midrule
                    
                    Nodes & 50 Icarus nodes \\
                    LTE connectivity & 8 nodes equipped with LTE dongles \\
                    LTE dongles &Huawei E392, Huawei E3272, and Huawei E3372\\
                    Topology & Grid topology with adjacent nodes separated by 1.2 meters\\
                    \bottomrule
                \end{tabular}
            \end{minipage}
            
            \begin{minipage}{.49\linewidth}
                \begin{tabular}{p{0.23\textwidth}|p{0.70\textwidth}}
                    \toprule
                    \multicolumn{2}{c}{\textbf{Icarus nodes details}}\\ 
                    \midrule
                        OS & Ubuntu 14.04.1 LTS for LTE nodes and Ubuntu 12.04.1 LTS for the remining nodes\\
                        CPU	& Intel® Core™ i7-2600 Processor, 8M Cache, at 3.40 GHz\\
                        RAM	& 8GiB DDR3\\
                        Wireless Interfaces	& Atheros 802.11a/b/g and  Atheros 802.11a/b/g/n (MIMO)\\
                    \bottomrule
                \end{tabular}
            \end{minipage}\\\\

            \begin{minipage}{\linewidth}
                \centering
                \begin{tabular}{p{0.22\textwidth}|p{0.40\textwidth}}
                    \toprule
                    \multicolumn{2}{c}{\textbf{University of Pisa cloud infrastructure details}}\\ 
                    \midrule
                    Guest OS & Ubuntu 18.04.3 LTS \\
                    Host CPU & Intel(R) Xeon(R) Gold 5120 CPU @ 2.20GHz \\
                    Number of allocated cores &2\\
                    Guest RAM & 4GiB\\
                    \bottomrule
                \end{tabular}
                
            \end{minipage}\\
            
        \end{tabular}
    \caption{The setup used to collect experimental data.}
    \label{Table:testbedcharacteristics}
    
\end{table*}

\subsection{Active measurements}
\label{subsection:Active-measurements}

\begin{figure}[t!]
    \centering
     \includegraphics[width=0.48\textwidth]{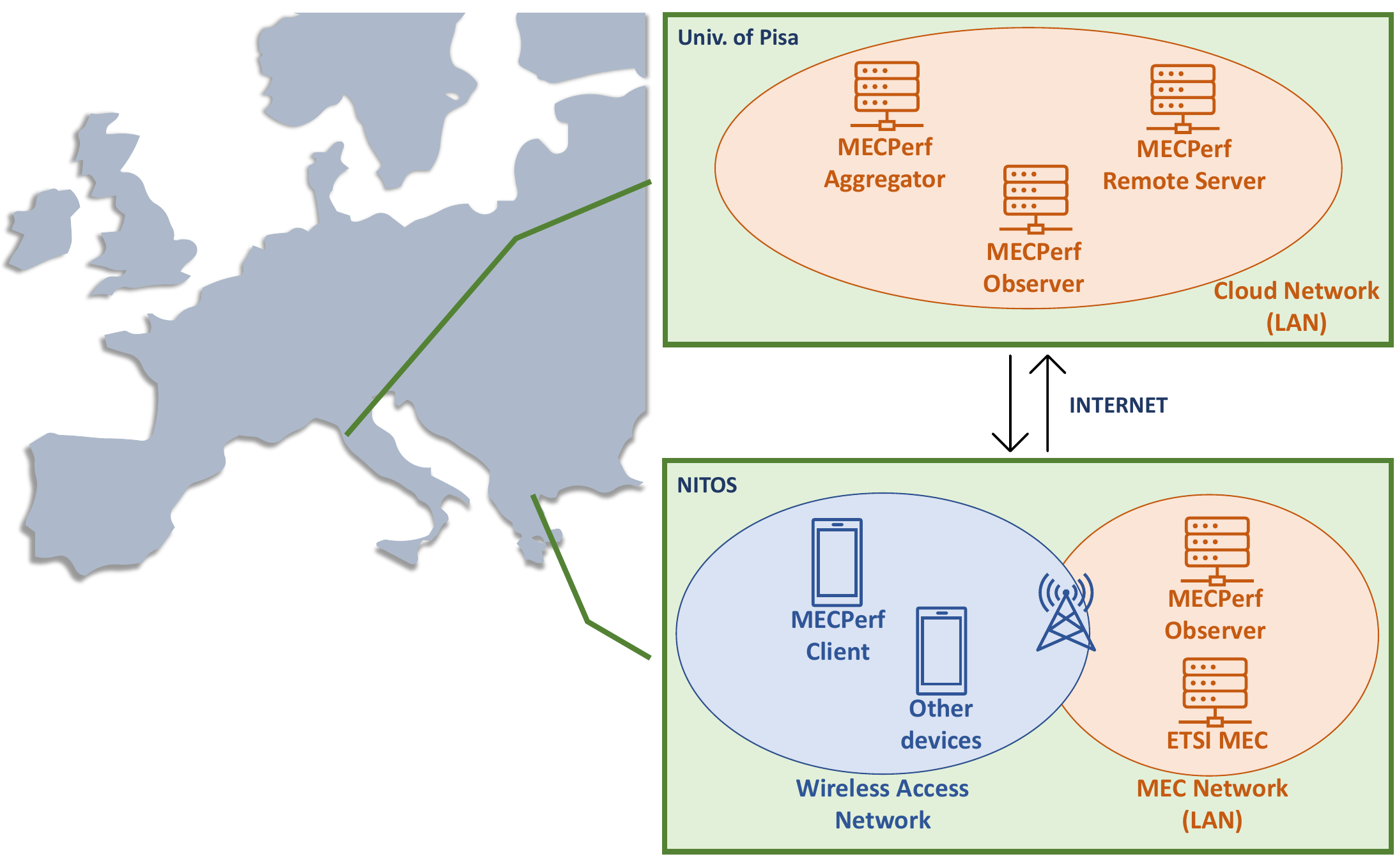}
    \caption{The setup used to collect active measurements.
    \label{fig:activeExperiments_setup}}
\end{figure}

\begin{figure*}[t!]
    \centering
    \subfloat[][Wi-Fi]{
        \label{fig:TCPBandwidthBoxplot_wifiDownstream}
        \includegraphics[width=0.47\textwidth]{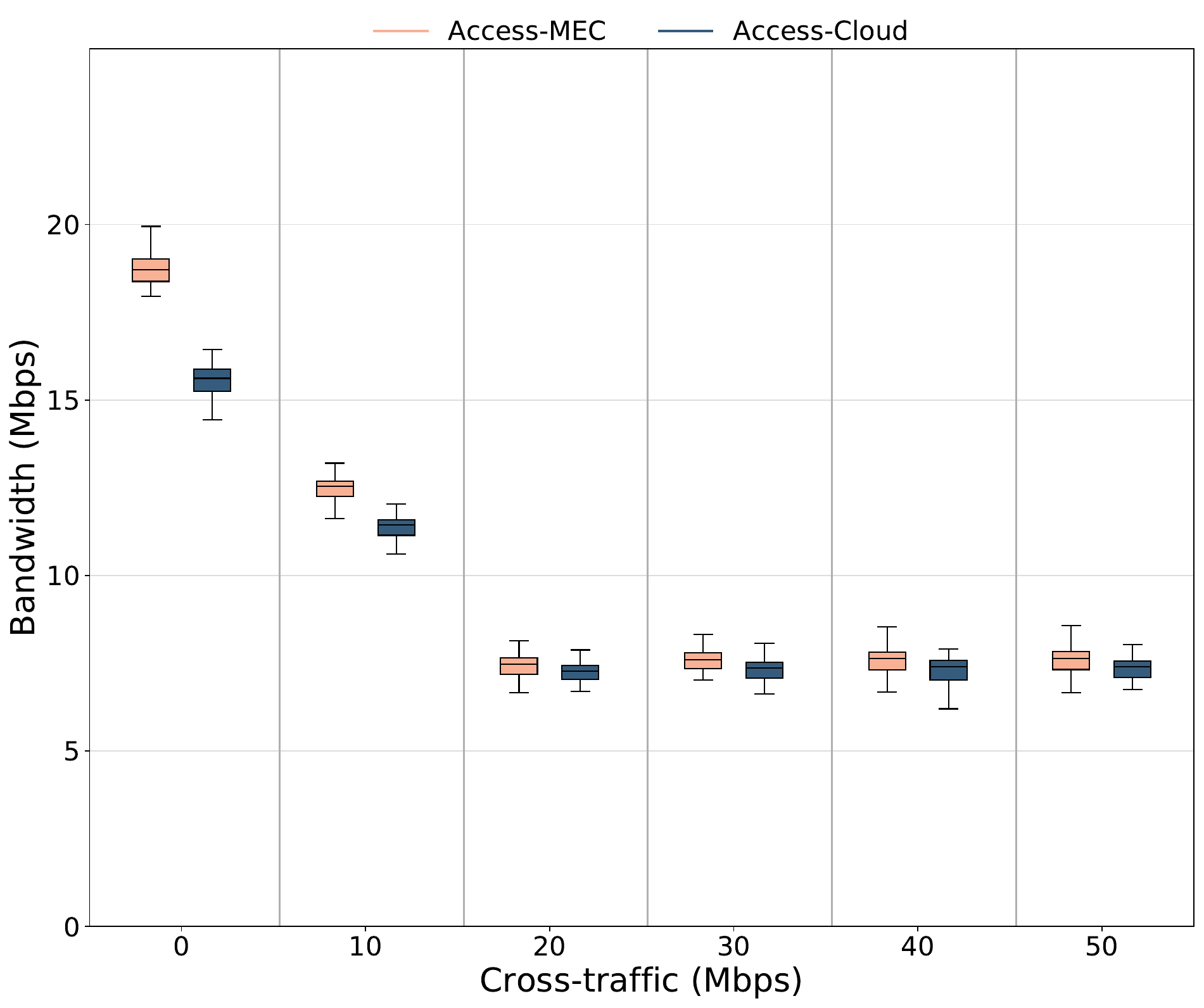}
        }
        \hfill
        \subfloat[][LTE]{
        \label{fig:TCPBandwidthBoxplot_lteDownstream}
        \includegraphics[width=0.47\textwidth]{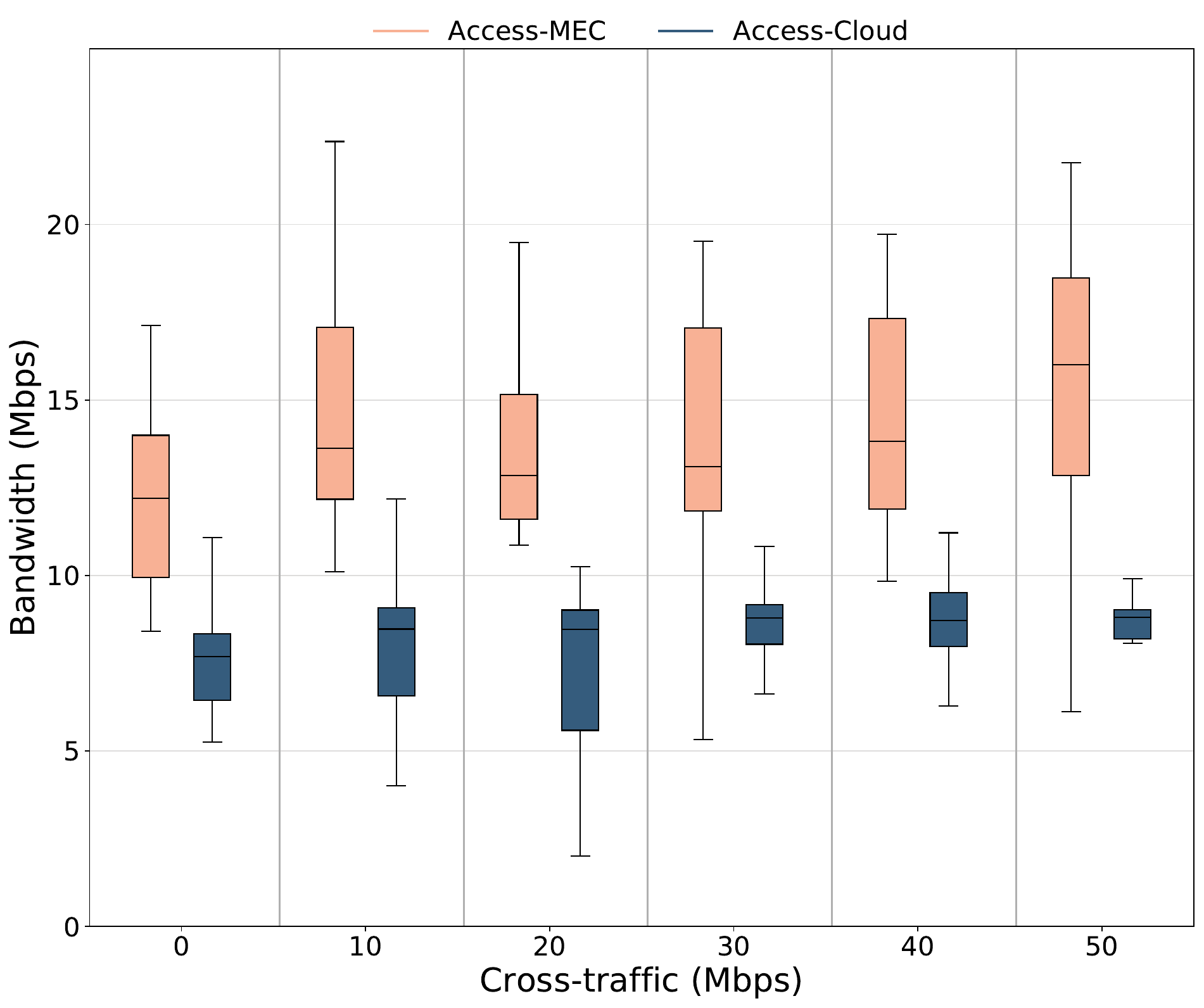}
        }
        
    \caption{TCP bandwidth (downstream) for Wi-Fi and LTE active experiments; the line within boxplots represents the median value, boxes show the Q1-Q3 range, whiskers include the values in 1.5 IQR.}
    \label{fig:TCPBandwidthBoxplot_Downstream}
\end{figure*}

The setup used for active experiments is depicted in Figure \ref{fig:activeExperiments_setup}. First, we built a wired MEC-enabled network and a wireless access network inside the NITOS testbed. The access network hosted a MECPerf Client and another device that was used to generate different levels of cross-traffic. The goal was to collect real-world data that can be representative of the experienced network performance when the access network is characterized by a non-negligible load, possibly imposed by other users. Cross-traffic was produced using iPerf3. We collected measurements using both Wi-Fi and LTE as access technologies.

Two MECPerf Observers were deployed as measuring end-points: the first was hosted inside the MEC network, and it was used to collect the performance metrics of the access-MEC segment, while the second was hosted ad the University of Pisa and it was used to collect the access-cloud segment performance metrics, i.e., corresponding to a scenario where the cloud is geographically far from the client. 
Initially, the MECPerf Client interacts with an ETSI MEC server to find the instance of MECPerf Observer to be involved in the data collection. Finally, a MECPerf Aggregator and a MECPerf Remote Server were hosted at the University of Pisa. The physical distance between the NITOS testbed and the remote server is compatible with the one of a client accessing the cloud in the core network (i.e., not in close proximity and involving transit on the public Internet).

We performed a set of experiments using both Wi-Fi and LTE access technologies. For LTE, we used Huawei E3272 modules, which have a maximum communication speed of 150 Mbps in download and 50 Mbps in upload. We repeated the experiment eight times a day at regular intervals to capture the possible effects due to external traffic, especially on the public Internet when communicating with the remote cloud. 
We also varied the amount of cross-traffic on the access network from 0 to 50 Mbps.
Each experiment included 10 upstream measurements and 10 downstream measurements of both bandwidth and latency. Measurements were based on both TCP and UDP. In detail, we used a stream of 1024 packets of 1420 bytes for each TCP bandwidth measurement, 25 packet pairs of 1420 bytes for the UDP capacity, and 25 RTT with transfers of 1 byte for both TCP and UDP latency. 

Figure \ref{fig:TCPBandwidthBoxplot_Downstream} shows the results for TCP bandwidth measurements. 
The bandwidth measured in the access-MEC segment is always higher than the bandwidth of the access-cloud segment. This is explained by the fact that the TCP protocol throughput is negatively influenced by the delay (the higher the delay, the lower the throughput). Similar results have also been obtained in the opposite direction. 
Wi-Fi and LTE react differently when cross-traffic is injected into the access network. For the Wi-Fi experiments, the access-MEC and the access-cloud TCP bandwidth values decrease as the cross-traffic increases. As the amount of cross-traffic gets larger, the difference between the bandwidth on the access-MEC segment and the bandwidth on the access-cloud segment becomes smaller. In particular, for cross-traffic rates higher than 20 Mbps, the difference becomes negligible. In other words, for the considered incarnation of the edge computing paradigm, a moderate competition with other users on the Wi-Fi access can be sufficient to vanish the benefits, in terms of bandwidth, brought by the reduced distance of the edge-based solution.
Results concerning bandwidth measurements when using LTE, show that
the performance seems to be marginally affected when the level of cross-traffic gets higher. Results are also characterized by increased variability. These somehow counter-intuitive results can be explained in different ways. First, the nominal capacity of LTE is higher than the one of Wi-Fi, so the amount of cross-traffic is less likely to produce a large, negative interference. Second, the different access mechanisms to the medium, as well as the scheduling mechanisms of LTE, tend to isolate the application traffic from the one artificially injected.
In addition, the difference between the two paths (access-MEC vs. access-cloud) remains significant even in the presence of a large amount of cross-traffic.
\subsubsection*{Takeaway} The relative networking performance induced by the different placement strategies (centralized vs. edge) is significantly affected by the adopted communication technology and by the load on the network. These factors are not adequately considered in most of the existing studies,  where the adoption of simplified network models is not fully representative of the complexity of the involved elements and their interaction.

\subsection{Passive and Self Measurements}

\begin{figure*}[!t]
    \centering
    \subfloat[][Self measurements, access-MEC segment.]{
        \label{subfig:selfmec}
        \includegraphics[width=0.48\textwidth]{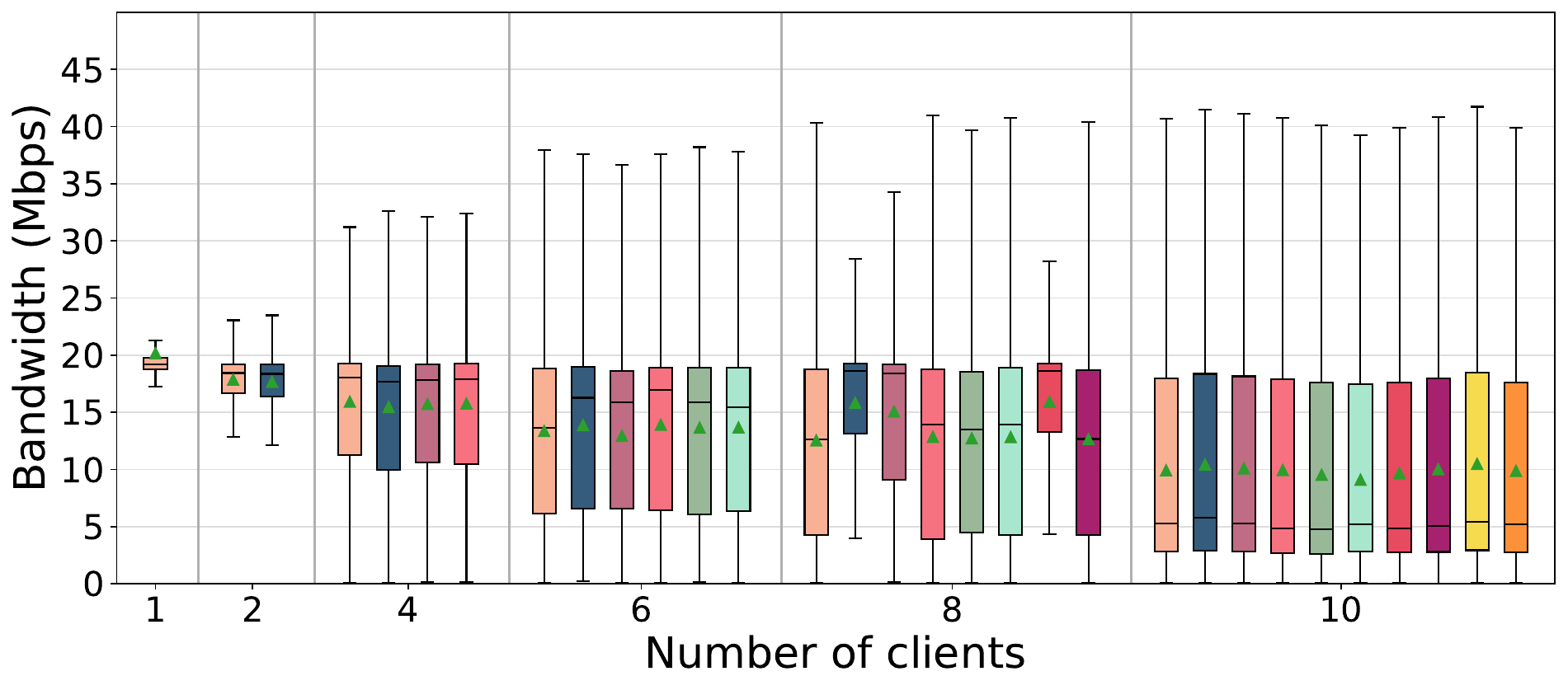}
    }
    \subfloat[][Self measurement, access-cloud segment.]{
        \label{subfig:selfcloud}
        \includegraphics[width=0.48\textwidth]{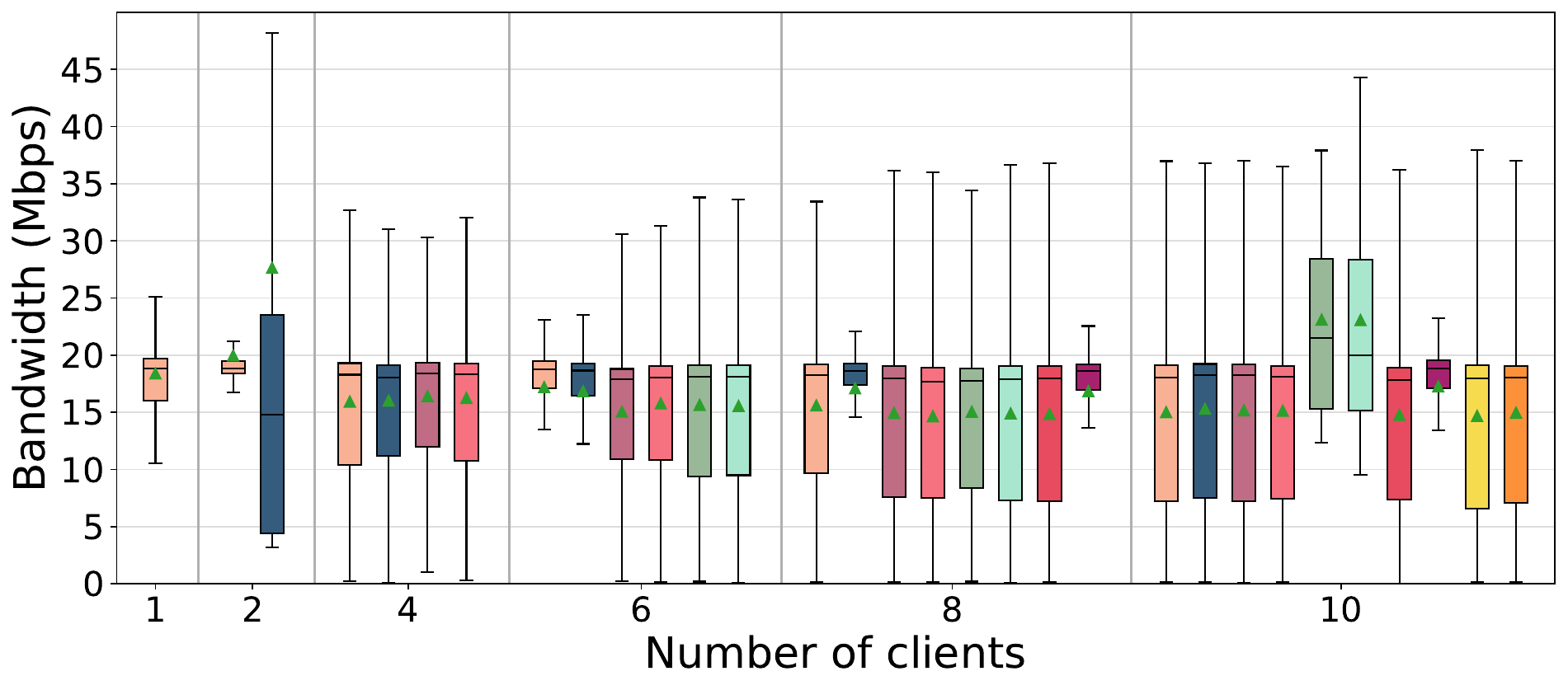}
    }\\
    \subfloat[][Passive measurements, access-MEC segment.]{
        \label{subfig:passivemec}
        \includegraphics[width=0.48\textwidth]{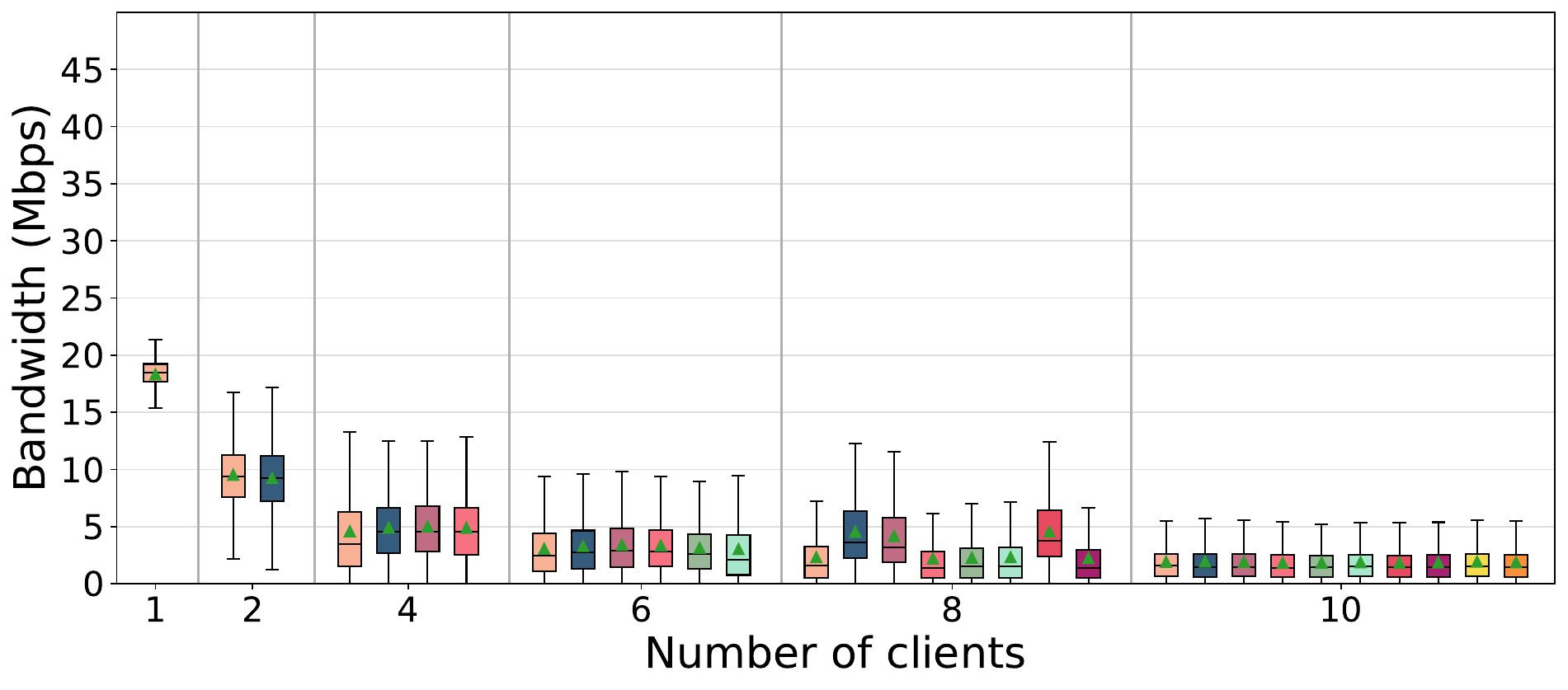}
    }
    \caption{Boxplots of the self and passive bandwidth measurements.}
    \label{fig:selfpassivecompare}
\end{figure*}

Passive and Self measurements have been performed together.
This happened for two reasons. First, the two approaches do not interfere with each other, thus we were able to optimize the execution times of the experiments. Second, running both approaches together allows us to perform a comparison between them, as highlighted below. Similarly to active measurements, we built a wireless access network and a MEC-enabled network inside NITOS. The application server was executed, alternatively, both in the NITOS's edge infrastructure and in the remote cloud infrastructure belonging to the University of Pisa. The considered application was video streaming, as video streaming is one of the six relevant use cases originally identified by ETSI for MEC \cite{hu2015mobile}. Our experiments were based on the Dynamic Adaptive Streaming over HTTP protocol (DASH)~\cite{6077864}. From 1 to 10 clients of a DASH video streaming application were positioned in the access network, together with the cross-traffic generator, all connected via Wi-Fi technology. The DASH streaming app client is able to measure the downlink bandwidth at which the video is downloaded from a server and communicates it to the MECPerf Aggregator. At the same time, the video streaming traffic is intercepted by the MECPerf Observer, positioned in the MEC-enabled network, and used to compute the KPIs according to the passive approach. KPIs were computed when downloading videos from two separate DASH servers, as mentioned, one positioned in the MEC network and the other in the remote cloud. Thus, the first is used to compute the access-MEC performance, and the other one the access-cloud performance. The DASH servers provide four videos with different lengths: 1, 5, 12, and 25 minutes. Each video is encoded in chunks at multiple bitrates: 200 Kbps, 500 Kbps, and 700 Kbps. The DASH video clients initially interact with an ETSI MEC server to obtain the address of the DASH server to be used. Then the clients start downloading video fragments at different bitrates, depending on the network conditions that they experience.

We first compare the results obtained via self measurements and via passive measurements. Figures~\ref{subfig:selfmec} and~\ref{subfig:passivemec} show an example of the results obtained by the two methods. For simplicity, we choose to show just the results for the access-MEC segment of the download of the 25 minutes video with no cross traffic, as the other configurations produced similar results. Since passive measurements return the number of bytes contained in each sniffed packet and the timestamp at which the packet has been captured, to compute the bandwidth, we divided the duration of a single run of the experiments into bins of 0.5 seconds. For each bin, we computed the average bandwidth. As can be observed, the performance computed via passive measurements is always lower than the one computed via self measurements. This highlights the importance of self measurements if one is interested in simulating applications that do not produce a constant stream of data at the application level. However, it must be noted that the passive measurement results are produced by aggregation, and every packet can be found in the raw measurement values, thus they can be more suitable for simulating a data stream of a certain application at the network level.

We then compare the results obtained by self measurements in the access-MEC and access-cloud segments, as shown in Figures~\ref{subfig:selfmec} and~\ref{subfig:selfcloud}. We observe that, when increasing the amount of traffic in the access network, e.g., with a high number of DASH clients, the bandwidth median values for the access-MEC segment start to decrease with respect to the ones of the access-cloud segment. We attribute this behavior to the different setup of the server machines hosting the DASH server. In particular, the machines located at the University of Pisa are more powerful and can better handle a high number of clients. This happens because when dealing with real-world devices not everything can be in control of the experimenter. While this aspect could be considered as an anomaly, we believe that it is instead evidence of the importance of deriving models and results from real systems.

\begin{figure*}[!t]
    \centering
    \subfloat[][Self measurements.]{
        \label{subfig:selfedgetime}
        \includegraphics[width=0.48\textwidth]{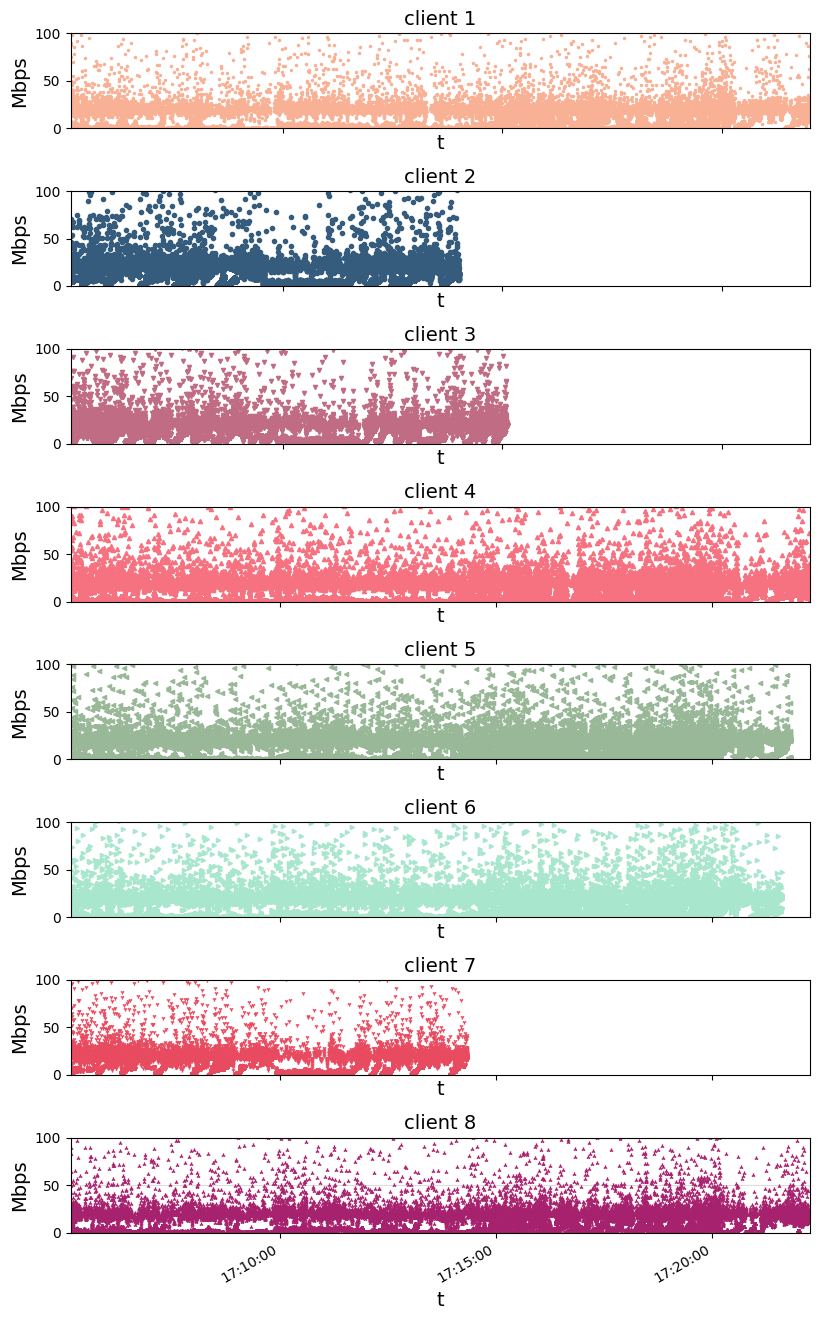}
    }
    \subfloat[][Passive measurements.]{
        \label{subfig:passiveedgetime}
        \includegraphics[width=0.48\textwidth]{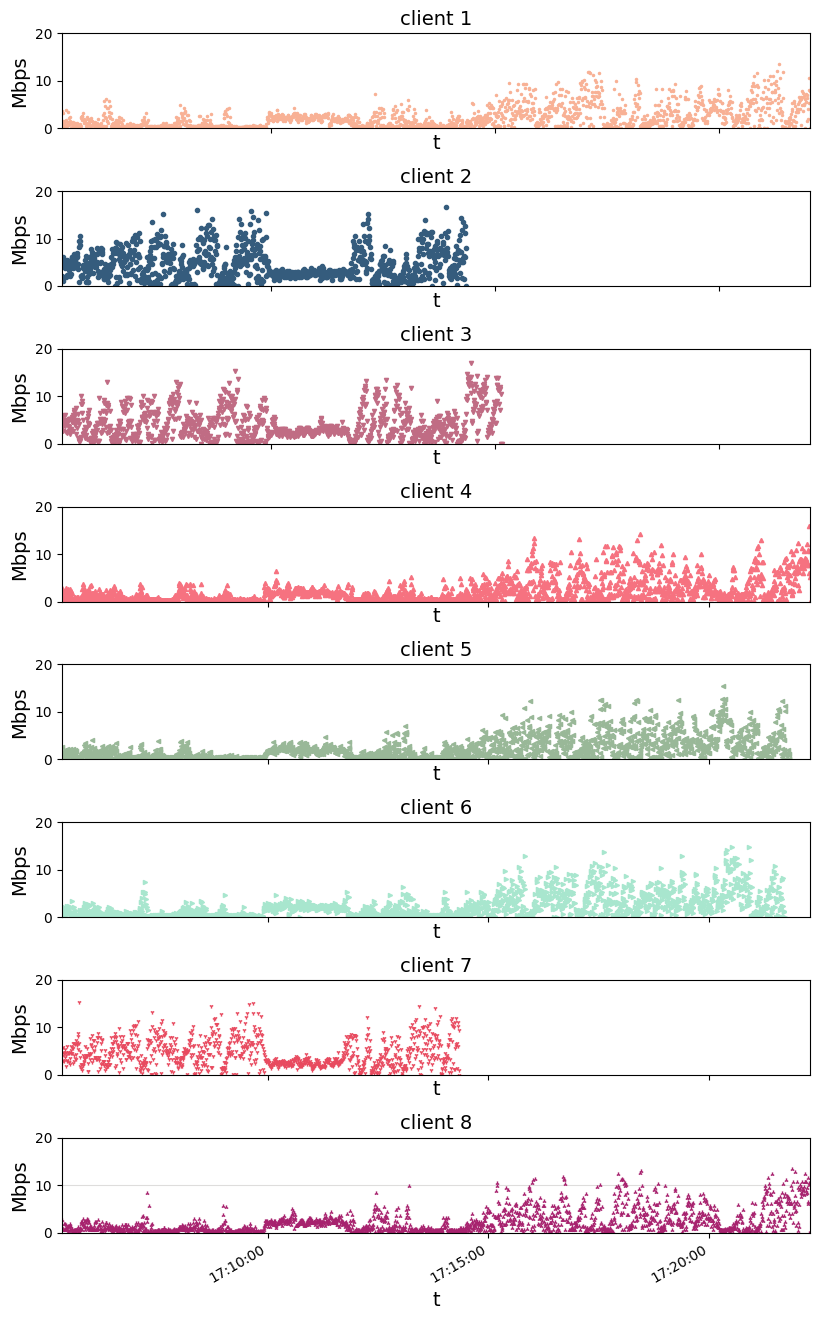}
    }
    \caption{Bandwidth values over time for self and passive measurements.}
    \label{fig:selfpassiveedgetime}
\end{figure*}

When multiple clients are involved, some of them can gain an advantage over the others, as can be seen in Figure~\ref{subfig:selfmec} when observing the results of the access-MEC segment for eight clients. We confirmed this behavior by analyzing the raw results obtained by each client. Figure~\ref{fig:selfpassiveedgetime} shows the bandwidth values plotted over the time for both self and passive measurements. The figure clearly shows that three out of eight clients complete the video transfer in approximately half the time compared to the others. In addition, the passive measurements show how the average bandwidth of the other clients is distinctly lower in the first half of the experiment and raises when the three ``fast'' clients complete their transfer. We believe that this behavior is caused by internal mechanisms of the DASH client, which can not be reproduced without a complex analysis of the client source code. This again highlights the importance of having real-world data for generating realistic simulations.

\begin{figure}[t!]
    \centering
     \includegraphics[width=0.48\textwidth]{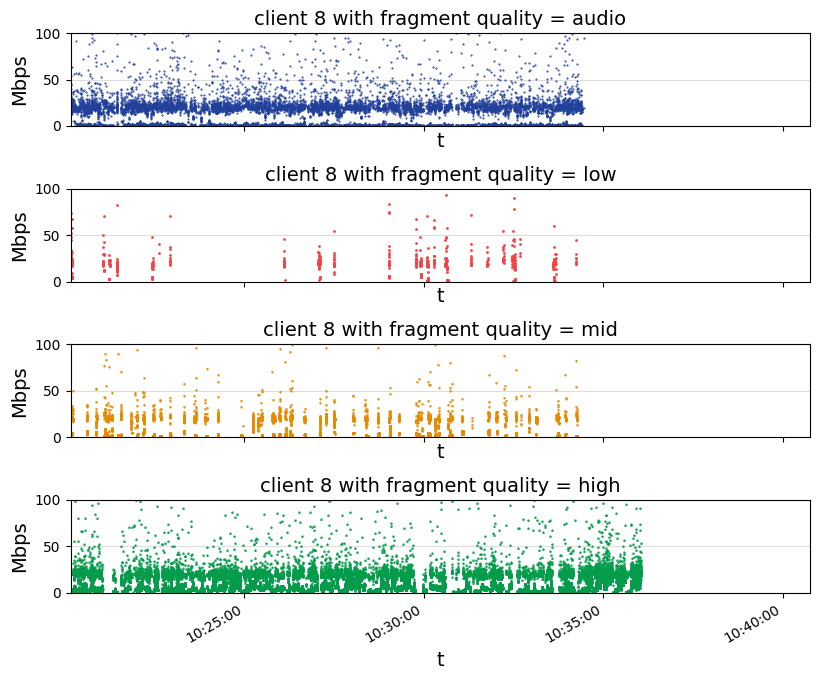}
    \caption{Bandwidth values over time for self measurements.
    \label{fig:selfchunks}}
\end{figure}

The DASH client uses multiple flows for downloading video chunks at different bitrates and audio chunks. MECPerf is able to identify the multiple flows and, via self measurements, also to tag the type of transfer according to the video quality or whether or not it is an audio chunk, as highlighted in Figure~\ref{fig:selfchunks}, which shows the bandwidth values measured via self measurements for the different types of chunks. As can be observed, there is some sort of layering in the values, which is particularly evident in the audio chunks. In addition, the chunks encoded at the different bitrates are downloaded according to a time-division behavior. We believe that these elements can again be exploited to enrich the data, which can be subsequently used to produce more realistic and detailed evaluations. For instance, real-world traces were used for understanding the different behaviors of adaptive video streaming algorithms~\cite{licciardello2020understanding}.

\subsubsection*{Takeaway} 

Active measurements are able to provide application-independent indicators that capture the performance of the networking infrastructure. Passive and self measurements provide a better view when the communication and execution schemes of an application are intertwined. In detail, self measurements provide a view of the edge network from the application side, whereas passive ones capture the communication patterns as seen from a bottom-up perspective. 
Despite performing similar operations and starting from macroscopically equal conditions, clients are characterized by performance levels that are not always homogeneous.

\section{A Case Study}
\label{sec:casestudy}

This section summarizes the different contributions of the paper within the use case of edge computing platform federation, as envisioned by \ac{GSMA} and briefly introduced below.
We have developed a toy simulator of two edge domains where MECPerf (see Section~\ref{sec:datacollection}) is used to collect real-time measurements of the network performance, which are exchanged between the two operators.
The simulator is bundled as an example with our software on GitHub, which uses the library described in Section~\ref{sec:mecperflib}, which in turn is driven by the dataset collected as illustrated in Section~\ref{sec:experiments}.

\noindent\textbf{Background.} In early 2020, the \ac{GSMA} has published a white paper~\cite{GsmaWhitePaper2020} that introduces the \textit{operator platform} concept, which is a first attempt to standardize in a top-down manner the federation of edge computing resources belonging to multiple operators.
This effort, which is still ongoing, complements the technical activities within \ac{ETSI} and \ac{3GPP} related to managing virtualized computation, storage, and communication resources within a single \ac{MNO}, and it is very significant since it comes from the biggest global association of telecom operators.
The operator platform concept envisions the definition of many interfaces among all the actors in an edge computing scenario: application provider (today's \ac{OTT} players), edge computing platform provider, end-users, and operators.
Quite understandably, the role of the operator is central, according to GSMA.
Since one of the most compelling reasons to use edge computing resources at all is that the user terminals can enjoy lower latency and higher throughput, we believe that real-time measurement of the network performance, as enabled by MECPerf, will be pivotal to adapt the system configuration to (possibly fast) changing conditions and provide the end-users with best possible \ac{QoE}.

\begin{figure}[!t]
    \centering
    \includegraphics[width=0.47\textwidth]{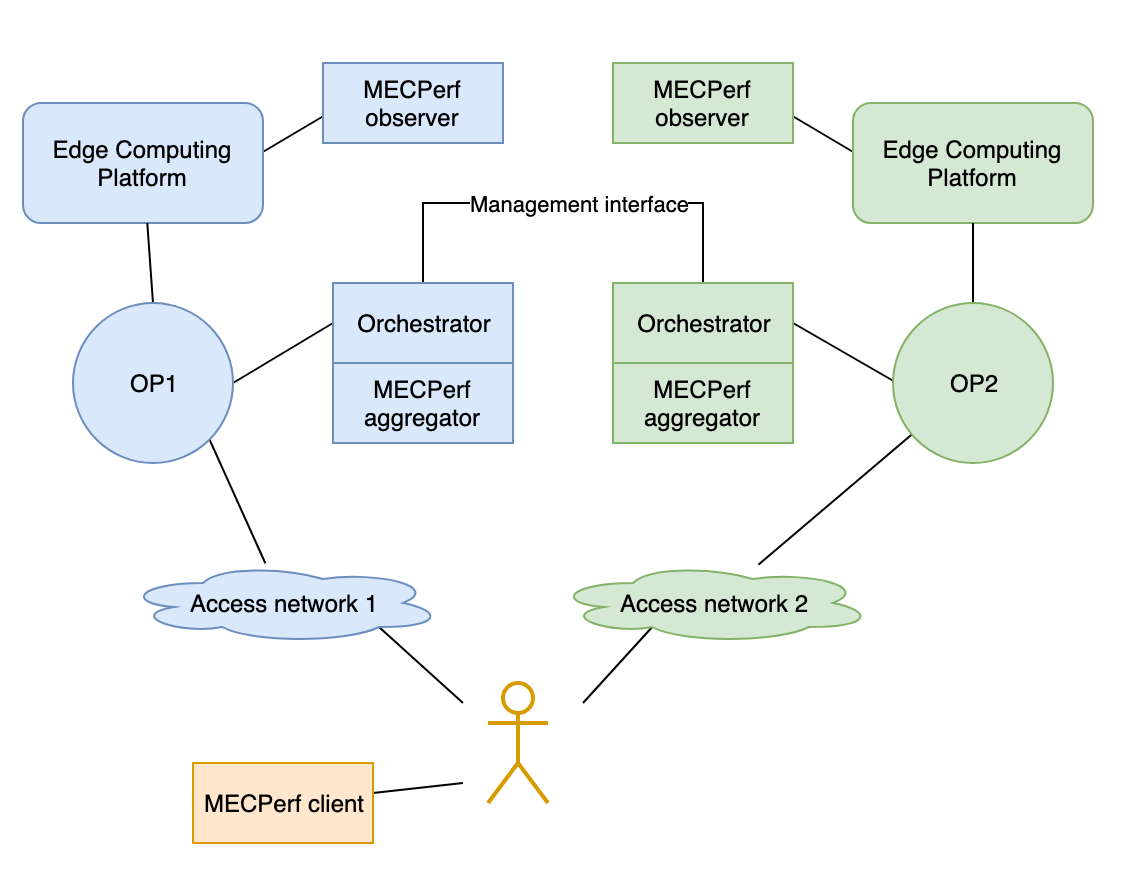}
    \caption{GSMA edge operator use case.}
    \label{fig:use-case}
\end{figure}

\noindent\textbf{Use case.} In Figure~\ref{fig:use-case} we show a scenario with two telco operators OP1 and OP2, both with edge computing resources.
We assume that the respective resource orchestrators are federated via a dedicated management interface, whose details are irrelevant to this study.
According to the architecture of MECPerf discussed so far, the logical location of its modules would be: MECPerf client integrated with the user terminal operated by the end-user, one or more MECPerf observers executing on edge computing platform resources, and a single MECPerf Aggregator, which can be co-located with the orchestrator since it provides the latter with useful information for dynamic system optimization.
For example, one optimization performed by both orchestrators could be the following: if the network performance of a given client using edge computing resources becomes too low, then migrate the client to the peer operator.

\noindent\textbf{Toy simulator.} The simulations are time-discrete and include 100 clients, all potentially connected to both OP1 and OP2 via their respective access networks, which have independent network characteristics.
Each client alternates active and inactive periods of random duration, both drawn an exponentially distributed random variables with a mean 10 time slots.
At the beginning of an active period, the client is associated with two network measurement traces, one per access network, and it subscribes to the edge computing services of an operator, picked at random.
Every network measurement trace can be LTE or Wi-Fi and have any possible cross-traffic amount from 0~Mbps to 50~Mbps, as available in the Fed4Fire+ MECPerf dataset provided. 
At the end of every time slot, each orchestrator collects RTT measures from the subscribed clients (= it queries the MECPerf Aggregator), then it forces the migration of a fraction of users $\gamma$ to the other operator, i.e., those with higher RTT values reported.
Such $\gamma$ is a system parameter that in practice would be subject to optimization: a higher $\gamma$ means a more aggressive stance towards migration of clients, which seeks to improve network performance at the cost of a higher rate of migrations, which are costly operations (such cost is not accounted for in our simplified simulation model).

\begin{figure}[!t]
    \centering
    \includegraphics[width=0.47\textwidth]{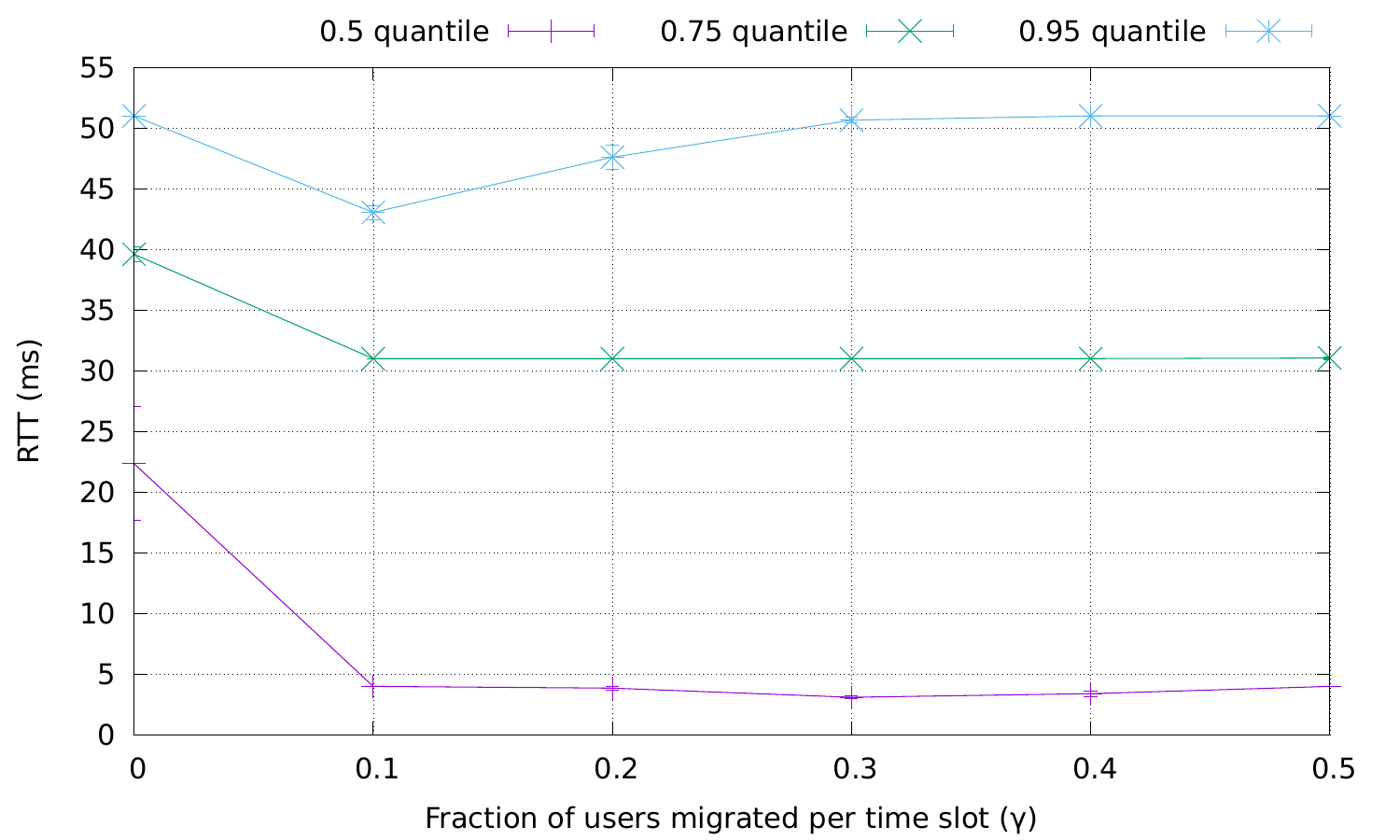}
    \caption{0.5, 0.75, 0.95 RTT quantiles with varying $\gamma$.}
    \label{fig:plot-rtt}
\end{figure}

\noindent\textbf{Results.} In Figure~\ref{fig:plot-rtt}, we report the 0.5, 0.75, and 0.95 quantiles of RTT when increasing $\gamma$ from 0 (i.e., no migration) to 0.5 (i.e., at every time slot half clients are migrated to the peer operator).
A 95\% confidence interval is also shown in the plot, measured across 20 independent replications of each scenario.
As can be seen, for all the quantiles, the trend is not monotone: migration helps in keeping a lower RTT, but as $\gamma$ increases too much, then the RTT quantiles increase again.
These simple simulations show that an optimum value of $\gamma$ exists around $0.1$ in terms of the RTT alone and that an overall optimum value also considering the migration rate can be found by an appropriate weighting of the two metrics.

\noindent\textbf{Wrap-up.} This section has served as a step-by-step guide from the identification of a use case relevant to MECPerf (i.e., edge domain federation) to the implementation of a simulator for performance evaluation (available as an example) and collection and analysis of results.
\section{Conclusion and Future Directions}
\label{sec:conclusion}

Edge computing has the potential to bring huge benefits in terms of network performance, from the reduced latency and increased bandwidth experienced by clients when accessing networked services to the improved locality of Internet traffic. However, to fully unlock the potential of edge computing, system designers need to quantitatively assess the impact of the different design alternatives or dynamically tune their runtime optimization strategies. 

A measurement-driven approach to design and runtime optimization can be effective in achieving such goals, especially in an edge computing scenario. In fact, the performance of edge-based systems is heavily influenced by a mix of computational and networking factors that is difficult to reproduce when facing the problem from a purely analytic or simulative perspective. Both MECPerf and the collected data are available to other researchers and practitioners interested in evaluating their edge computing infrastructure or improving existing simulators and/or models. We also believe that this paper contributes to the landscape of edge computing literature by providing a view that is not limited to a tool or to the results originated from a data collection. The importance of the methodology that we presented by describing the collection tool, the data collection phase, and the results of the experiments is demonstrated by the case study, where the most relevant parameter of operation of the evaluated migration strategy is tuned through a measurement-driven approach.

As future work, we would like to integrate MECPerf with the MEC Orchestrator, which is responsible for managing the MEC applications, e.g., by selecting the proper target MEC server, while taking into account the requirements of the applications themselves (in terms of throughput and latency). Finally, MECPerf is currently able to provide information about the current status of the network and about its past. We would like to add forecasting capabilities, even limited to a short time-frame, as information about the future of the network can be extremely valuable for managing components and applications. Techniques like the one described in \cite{10.1145/3419394.3423643} could be adapted to steer the prediction as soon as new measurements are collected.

\section*{Acknowledgment}

This work is partially funded by the Horizon 2020 Research
and Innovation Programme under grant agreement No. 732638
(Fed4Fire+) through the 6th open call, and by the Italian
Ministry of Education and Research (MIUR) in the framework
of the CrossLab project (Departments of Excellence). The
views expressed are solely those of the authors.

\bibliographystyle{elsarticle-num}
\bibliography{mecperf}

\end{document}